\documentclass[onecolumn]{IEEEtran}

\usepackage[utf8]{inputenc} 
\usepackage[T1]{fontenc}
\usepackage{url}
\usepackage{ifthen}
\usepackage{cite}
\usepackage[normalem]{ulem}
\usepackage[cmex10]{amsmath} 

\usepackage{amsfonts}
\usepackage{amssymb}
\usepackage{amsthm}
\usepackage{algorithmicx}
\usepackage{algpseudocode}
\usepackage{algorithm}
\usepackage{graphicx}
\usepackage{cite}
\usepackage{calc}
\usepackage{color}
\usepackage{epsfig}
\usepackage{setspace}
\usepackage{multirow}
\usepackage{mathtools, cuted}
\usepackage{epstopdf}
\usepackage{lipsum}
\usepackage{mathtools}
\usepackage{cuted}
\usepackage{caption}
\usepackage{placeins}
\usepackage{subcaption} 
\newtheorem{defn}{Definition}
\newtheorem{thm}{{\cal T}heorem}
\newtheorem{cor}{Corollary}
\newtheorem{prop}{Proposition}
\newtheorem{lem}{Lemma}
\newtheorem{conj}{Conjecture}
\newtheorem{constr}{Construction}
\newtheorem{note}{Remark}
\newtheorem{example}{Example}
\newcommand{\bit}{\begin{itemize}}
	\newcommand{\eit}{\end{itemize}}
\newcommand{\bcor}{\begin{cor}}
	\newcommand{\ecor}{\end{cor}}
\newcommand{\beq}{\begin{equation}}
	\newcommand{\eeq}{\end{equation}}
\newcommand{\beqn}{\begin{equation}}
	\newcommand{\eeqn}{\end{equation}}
\newcommand{\bea}{\begin{eqnarray}}
	\newcommand{\eea}{\end{eqnarray}}
\newcommand{\bean}{\begin{eqnarray*}}
	\newcommand{\eean}{\end{eqnarray*}}
\newcommand{\ben}{\begin{enumerate}}
	\newcommand{\een}{\end{enumerate}}
\newcommand{\bdefn}{\begin{defn}}
	\newcommand{\edefn}{\end{defn}}
\newcommand{\bnote}{\begin{note}}
	\newcommand{\enote}{\end{note}}
\newcommand{\bprop}{\begin{prop}}
	\newcommand{\eprop}{\end{prop}}
\newcommand{\blem}{\begin{lem}}
	\newcommand{\elem}{\end{lem}}
\newcommand{\bthm}{\begin{thm}}
	\newcommand{\ethm}{\end{thm}}
\newcommand{\bconj}{\begin{conj}}
	\newcommand{\econj}{\end{conj}}
\newcommand{\bconstr}{\begin{constr}}
	\newcommand{\econstr}{\end{constr}}
\newcommand{\bpf}{\begin{proof}}
	\newcommand{\epf}{\end{proof}}

\begin{document}
\title{On MDS Convertible Codes \\ in the Merge Regime} 



\author{%
	\IEEEauthorblockN{
		 Vinayak Ramkumar,
		Xiangliang Kong,
        G. Yeswanth Sai,
        		Myna Vajha, 
                 M. Nikhil Krishnan
	}		
    \thanks{This work was presented in part at the 2025 IEEE International Symposium on Information Theory (ISIT) \cite{Convertible_ISIT25}. 

    Vinayak Ramkumar is with the Institute for Communications Engineering, Technical University of Munich, Germany.
    Xiangliang Kong is with the Department of Electrical Engineering--Systems, Tel Aviv University, Israel.
    G. Yeswanth Sai is with the Mehta Family School of Data Science \& Artificial Intelligence, Indian Institute of Technology Palakkad, India. Myna Vajha is with the Department of Electrical Engineering, Indian Institute of Technology Hyderabad, India. 
    M. Nikhil Krishnan is jointly with the Mehta Family School of Data Science \& Artificial Intelligence and the Department of Electrical Engineering, Indian Institute of Technology Palakkad, India. Emails: \{vinram93, rongxlkong, ysai0158, mynaramana, nikhilkrishnan.m\}@gmail.com
	}}

\maketitle

\begin{abstract}

	In large-scale distributed storage systems, erasure coding is employed to ensure reliability against disk failures. Recent work by Kadekodi et al. demonstrates that adapting code parameters to varying disk failure rates can lead to significant storage savings without compromising reliability. Such adaptations, known as \emph{code conversions}, motivate the design of \emph{convertible codes}, which enable efficient transformations between codes of different parameters.
    In this work, we study the setting in which  $\lambda$  codewords of an initial $[n^I = k^I + r^I,\, k^I]$ MDS code are merged into a single codeword of a final $[n^F = \lambda k^I + r^F,\, k^F = \lambda k^I]$ MDS code. We begin by presenting three constructions that achieve optimal \emph{access cost}, defined as the total number of disks accessed during the conversion process. The first two constructions apply when $\lambda \leq r^I$ and impose specific divisibility conditions on $r^I$ and the field size $q$. These schemes minimize both the per-symbol and the overall access cost. The third construction, which builds on a prior scheme by Kong, achieves minimal access cost while supporting arbitrary parameter regimes. All three constructions require field sizes that are linear in the final code length, and notably, the third construction achieves a field size that matches the lower bound implied by the MDS conjecture in almost all cases. In addition, we propose a construction that optimizes the \emph{bandwidth cost}, defined as the total number of symbols transmitted during conversion. This scheme is a refinement of Maturana and Rashmi’s bandwidth-optimal construction based on the piggybacking framework, and achieves reduced sub-packetization.
\end{abstract}
\begin{IEEEkeywords}
	Distributed storage, convertible codes, maximum distance separable codes, access cost, bandwidth cost 
\end{IEEEkeywords}

\section{Introduction}
Erasure codes are widely used in distributed storage systems to efficiently provide fault tolerance. An $[n, k]$ erasure code over a finite field $\mathbb{F}_q$ encodes a file by first dividing it into $k$ data symbols and then encoding them into $n$ coded symbols. Each coded symbol is stored on a separate disk to ensure resilience against disk failures. Nowadays, most storage systems employ a static erasure code, where the code parameters $n,k$
(typically chosen based on the expected disk failure rate) remain fixed throughout the system’s lifetime. However, this design may render the storage system either storage-inefficient or increasingly vulnerable to failures as the disks age, since the failure rates of storage devices in large-scale systems can vary significantly over time~\cite{BGPS07,PWB07,SDG10,MTDCLSCH15}. In~\cite{KadekodiRashmiGanger_FAST19}, it was shown that adapting code parameters to varying disk failure rates can improve storage efficiency. However, subsequent work~\cite{Kadekodi_osdi20} demonstrated that such adaptations may incur significant I/O overhead and adversely affect overall cluster performance. 

Motivated by the need to adapt code parameters to varying disk failure rates while maintaining low I/O overhead, Maturana and Rashmi introduced the convertible codes framework and initiated the theoretical study of code conversions in distributed storage systems~\cite{Maturana_Rashmi_TIT}. An $(n^I, k^I; n^F, k^F)$ convertible code consists of an initial $[n^I = k^I + r^I,\, k^I]$ code $\mathcal{C}^I$ and a final $[n^F = k^F + r^F,\, k^F]$ code $\mathcal{C}^F$. Given $M \triangleq \mathrm{lcm}(k^I, k^F)$, a set of $\lambda^I \triangleq M/k^I$ codewords from $\mathcal{C}^I$ are transformed into $\lambda^F \triangleq M/k^F$ codewords in $\mathcal{C}^F$ to represent the same batch of data through a conversion procedure. The efficiency of such conversions can be measured in terms of the following two metrics:
\begin{itemize}
    \item[(a)] \textit{Access cost}~\cite{Maturana_Rashmi_TIT} — the total number of disks accessed during the conversion.
    \item[(b)] \textit{Bandwidth cost}~\cite{MaturanaRashmi_BWMerge_TIT} — the total number of symbols transferred across the network during the conversion.
\end{itemize}

Due to the widespread applications of maximum distance separable (MDS) codes in modern storage systems, most existing works in the literature focus on the scenario where both the initial and final codes are linear MDS codes. The study of code conversions among MDS codes typically considers two regimes: the \emph{merge regime} and the \emph{split regime}. In the merge regime, where $\lambda = \lambda^I$ initial codewords are merged into a single final codeword,~\cite{Maturana_Rashmi_TIT} derives the following tight lower bound on the access cost:
\[
C_\text{access} = 
\begin{cases} 
	(\lambda + 1)r^F, & \text{if } r^F \leq \min\{r^I, k^I\}, \\
	\lambda k^I + r^F, & \text{otherwise}.
\end{cases}
\]
Note that the default conversion scheme, which accesses $k^I$ disks to read the information symbols of each initial codeword and writes new parity symbols of the final code to $r^F$ disks, incurs an access cost of $\lambda k^I + r^F$. Therefore, according to the bound above, the most interesting case arises when $r^F \leq \min\{r^I, k^I\}$. In~\cite{Maturana_Rashmi_TIT}, Maturana and Rashmi present an access-optimal construction in which both the initial and final codes are MDS codes with systematic generator matrices. The parity portions of these generator matrices are carefully designed Vandermonde matrices to ensure access cost optimality. However, this construction requires a field size of $q \ge \max\{2^{O(n^{F^3})},\, n^I - 1\}$, which becomes impractical when $n^F$ is large. Furthermore, Maturana and Rashmi also provide three families of access-optimal MDS convertible code constructions based on superregular Hankel arrays (see Table~\ref{tab:constr}). Although these constructions reduce the field size requirement compared to the aforementioned Vandermonde-based construction, they do not cover all parameter regimes where $r^F \leq r^I$.

Recently,~\cite{SaranshMaturanaRashmi} reduced the field size requirement for the construction using Vandermonde-matrix-based generator matrices to $O((k^F)^{r^F})$ in the general case. In~\cite{Kong}, Kong proposed a construction of access-optimal MDS convertible codes using a polynomial-based framework, in which both the initial and final codes are Generalized Reed–Solomon (GRS) codes. The conversion procedure in~\cite{Kong} is described as a transformation from $\lambda$ polynomials of degree less than $k^I$ into a single polynomial of degree less than $\lambda k^I$. The field size required for this construction is $q \ge (\lambda + 1)\max\{k^I, r^I\} + 1$, under the additional constraint that $\max\{k^I, r^I\} \mid (q - 1)$. 

Besides the aforementioned works on code conversions among MDS codes in the merge regime, there are also studies that focus on conversions in other regimes and for different classes of codes. For example, in~\cite{MaturanaMukkaRashmi_ISIT20}, the authors provide a construction of MDS convertible codes along with a matching lower bound on the access cost for the split regime, where $k^I = \lambda k^F$ and each initial codeword is split into $\lambda = \lambda^F$ final codewords. Moreover, they derive lower bounds on the access cost for general parameters $\{n^I, k^I, n^F, k^F\}$ and present corresponding MDS convertible code constructions that achieve them. Code conversions among locally recoverable codes are explored in~\cite{MaturanaRashmi_LRC_ISIT23, Kong}. However, the access-optimality results for these codes are restricted to the merge regime~\cite{Kong}.



In \cite{MaturanaRashmi_BWMerge_TIT}, Maturana and Rashmi initiated the study of optimizing bandwidth cost of conversion, i.e., the amount of data transfer required during the conversion process. For optimizing access cost, it suffices to explore scalar codes, i.e., codes for which each code symbol is an element of a finite field $\mathbb{F}_q$.  However, allowing partial downloads from disks can lead to lower bandwidth costs. 
Motivated by the literature on regenerating codes \cite{DimakisGWWR10, RashmiSK11, TamoWB13, YeB17, VajhaRPKLSKBYNH18}, the authors of \cite{MaturanaRashmi_BWMerge_TIT} explored the use of vector codes, where each code symbol is a vector in $\mathbb{F}_q^\alpha$. The parameter $\alpha \geq 1$ is referred to as {\it sub-packetization} in the literature.
Having $\alpha > 1$ enables the code conversion procedure to read only a fraction of the symbols (over $\mathbb{F}_q$) from each vector code symbol (over $\mathbb{F}_q^\alpha$), potentially reducing the bandwidth. The optimal bandwidth cost for vector MDS codes under the merge regime was characterized in \cite{MaturanaRashmi_BWMerge_TIT}, by providing lower bounds and matching constructions. This optimal bandwidth cost is given by: 
\beq C_{\text{bandwidth}} = 
\label{eq:bandwidth}
\begin{cases}
	\lambda \alpha \min\{k^I,r^F\}+r^F\alpha, & \text{if } r^I \geq r^F \text{or }k^I\leq r^F,\\
	\lambda \alpha (r^I+k^I(1-\frac{r^I}{r^F}))+r^F\alpha, & \text{otherwise.}
\end{cases}
\eeq 

According to the bound in equation~\eqref{eq:bandwidth}, the default conversion approach is already bandwidth-optimal when $k^I \leq r^F$. Moreover, in the case where $r^I \geq r^F$, one can easily verify that the optimal bandwidth is achievable using access-optimal MDS convertible codes. Therefore, the only non-trivial regime remaining is when $k^I > r^F > r^I$. For this regime, Maturana and Rashmi~\cite{MaturanaRashmi_BWMerge_TIT} proposed a bandwidth-optimal convertible code construction by \emph{piggybacking}\footnote{Piggybacking, introduced in~\cite{RashmiSR17_piggyback}, is a method for constructing vector codes from scalar codes.} access-optimal (scalar) MDS convertible codes. This piggybacking-based construction requires sub-packetization level $\alpha = r^F$. Notably, characterizing the minimum sub-packetization required for bandwidth-optimal vector MDS codes was posed as a challenging open problem in~\cite{MaturanaRashmi_BWMerge_TIT}. For the split regime, bandwidth costs were further studied in~\cite{MaturanaRashmi_BWSplit_ISIT}, and optimal constructions were presented for certain parameter settings.

\begin{table*}[!t]
	\vspace{0.1in}
	\begin{center}
		\scalebox{0.8}{
			\begin{tabular}{||p{1.7in}|p{2.5in}|p{1.6in}|p{0.6in}||} \hline 
				Construction & Supported Parameters & Field Size & Per-Symbol Access-Optimal?  \\ \hline
				Vandermonde-based \cite{Maturana_Rashmi_TIT} & All parameters& $\max\{2^{O((n^{F})^3)}, n^I-1\}$ & Y \\ \hline
				Vandermonde-based \cite{SaranshMaturanaRashmi} & All parameters& $O((k^F)^{r^F})$ & Y \\ \hline				 
				Hankel 1 \cite{Maturana_Rashmi_TIT} & $r^F \le \lfloor  \frac{r^I}{\lambda} \rfloor$ & $q \ge \max\{n^F-1, n^I-1\}$& Y \\ \hline
				Hankel 2  \cite{Maturana_Rashmi_TIT}& $r^F \le r^I-  \lambda +1$ &  $q \ge k^Ir^I$& Y\\ \hline
				Hankel 3  \cite{Maturana_Rashmi_TIT}& $r^F \le (s-\lambda+1)\lfloor \frac{r^I}{s}\rfloor+\max\{(r^I \mod s)-\lambda+1, 0\}$, $s \in \{\lambda, \lambda+1,
            \dots, r^I\}$ & $q \ge \max\{k^Is+\lfloor r^I/s\rfloor-1, n^I-1\}$ & Y \\ \hline
				Polynomial-based \cite{Kong}& All parameters &  $q \ge (\lambda + 1) \max\{r^I, k^I\}+1$, $\max\{r^I, k^I\} \mid (q-1)$ & N \\ \hline
				\multirow{2}{1.2in}{\it Sub-group-based $\mathit{1}$ (multiplicative)}
				& $\lambda \le r^I$ & $q \ge (k^I+1) r^I+1$, $r^I | (q-1)$ & Y \\ \cline{2-4}
				& $\lambda \le r^I-1$ & $q \ge (k^I+1) (r^I-1)+1$, $(r^I-1) | (q-1)$ & Y \\ \cline{2-4}
				& $\lambda \le r^I-2$ & $q \ge (k^I+1) (r^I-2)+1$, $(r^I-2) | (q-1)$ & Y \\ \hline
				\multirow{2}{1.2in}{\it Sub-group-based $\mathit{2}$ (additive)} & $\lambda \le r^I$ & $q \ge (k^I+1) r^I$, $r^I | q$ & Y \\ \cline{2-4}
				& $\lambda \le r^I-1$ & $q \ge (k^I+1) (r^I-1)$, $(r^I-1) | q$ & Y \\ \hline
				{\it Modified Polynomial-based} & All parameters & $  q \ge \max \{n^F-1, n^I-1\}$ & N \\ 
				\hline
			\end{tabular}
		}
		\caption{A summary of $(n^I = k^I+r^I, k^I; n^F = \lambda k^I+r^F, k^F = \lambda k^I)$ access-optimal MDS convertible codes in the merge regime for $r^F \leq \min\{r^I, k^I\}$. The constructions introduced in this paper are italicized. \label{tab:constr}}
	\end{center}
\end{table*}

\subsection{Our Contributions}
In line with previous works \cite{Maturana_Rashmi_TIT,SaranshMaturanaRashmi,Kong,MaturanaRashmi_BWMerge_TIT}, we focus on constructing MDS convertible codes with optimal access cost and optimal bandwidth cost in the merge regime. Specifically, our contributions are twofold:
\begin{itemize}
    \item [(1)] We present three low-field-size constructions of access-optimal MDS convertible codes for the merge regime. The first two constructions are \emph{subgroup-based constructions}, where the initial and final codes are described via systematic generator matrices with parity portions formed by Cauchy matrices using elements from specific subgroups of either $\mathbb{F}_q^*$ or $\mathbb{F}_q$. These constructions support a broader parameter range than the Hankel-array-based approaches in \cite{Maturana_Rashmi_TIT}. Furthermore, the resulting MDS convertible codes are not just access-optimal but also \emph{per-symbol access-optimal} (Definition~\ref{def per-sym-opt}, Section~\ref{sec:ao}). Our third construction is a modification of the polynomial-based approach from \cite{Kong}, which works without any parameter restrictions on initial and final codes. Notably, all three constructions require field sizes linear in the code lengths of both the initial and final codes, and in particular, the third construction can also achieve the minimal field size allowed by the MDS conjecture for almost all parameters.
    
    \item[(2)] Our second contribution is a construction of bandwidth-optimal vector MDS convertible codes in the merge regime, built on the piggybacking approach proposed in~\cite{MaturanaRashmi_BWMerge_TIT}. Compared to their construction, which requires a sub-packetization level of $\alpha = r^F$, our construction reduces this requirement to $\alpha = r^F/\gcd(r^F,r^I)$ while preserving bandwidth optimality. Moreover, we prove that $\alpha = r^F/\gcd(r^F,r^I)$ is in fact the minimal possible sub-packetization under certain natural assumptions. 
The sub-packetization reduction is particularly interesting from a practical standpoint. Lower sub-packetization enables easier implementation and has been an active research area in distributed storage; e.g., see~\cite{RashmiSR17_piggyback, RawTamGur_epsilonMSR1, LiLiuTang, VajhaRPKLSKBYNH18, eps_MSR_anyd, FnT}. It offers several benefits--such as bandwidth-efficient on-the-fly repair of missing partial blocks during degraded reads \cite{degradedReads,RawTamGur_epsilonMSR1}, reduced update complexity, and improved parallelism in encoding and decoding.

\end{itemize}

\subsection{Notation}

We use $\mathbb{Z}$ to denote the set of all integers and $\mathbb{N}$ to denote the set of all natural numbers, i.e., $\mathbb{N} \triangleq \{1, 2, \dots\}$. The interval $[a:b]$ is defined as $[a:b] \triangleq \{i \in \mathbb{Z} \mid a \leq i \leq b\}$. When $a = 1$, we use the shorthand $[b] \triangleq [1:b]$. Let $\mathbf{A}$ be an $m \times n$ matrix. The rows of $\mathbf{A}$ are indexed by $[0:m-1]$ and the columns by $[0:n-1]$. The entry at the intersection of the $i$-th row and $j$-th column is denoted by $\mathbf{A}(i,j)$. For $R \subseteq [0:m-1]$ and $C \subseteq [0:n-1]$, the submatrices obtained by restricting $\mathbf{A}$ to rows in $R$, columns in $C$, or both, are denoted by $\mathbf{A}(R,:)$, $\mathbf{A}(:,C)$, and $\mathbf{A}(R,C)$, respectively. We denote the finite field with $q$ elements by $\mathbb{F}_q$ and the identity matrix of size $a \times a$ by $\mathbf{I}_a$. A linear code of dimension $k$ and block length $n$ is referred to as an $[n,k]$ code. We use underbars to denote vectors, e.g., $\underline{x}, \underline{y}$, etc. 
All vectors in this paper are assumed to be row vectors, unless stated otherwise.
For a set $\mathcal{S}$, we use $|\mathcal{S}|$ to denote its cardinality. For a subset $\mathcal{S} \subseteq \mathbb{F}_q$ and an element $\beta \in \mathbb{F}_q$, we define $\beta \mathcal{S} \triangleq \{\beta s \mid s \in \mathcal{S}\}$ and $\beta + \mathcal{S} \triangleq \{\beta + s \mid s \in \mathcal{S}\}$. The empty set is denoted by $\emptyset$. For $a, b \in \mathbb{Z}$, we use $a \mid b$ to indicate that $a$ divides $b$.


\subsection{Organization of the Paper} 

The rest of this paper is organized as follows. Section~\ref{sec:bg} introduces the formal definition of convertible codes in the merge regime, along with preliminary concepts and results needed for subsequent constructions and proofs. Section~\ref{sec:sg_constr} presents our two subgroup-based constructions of access-optimal MDS convertible codes. In Section~\ref{sec:mod_poly_constr}, we describe the modified polynomial-based construction of access-optimal MDS convertible codes. Our bandwidth-optimal code construction appears in Section~\ref{sec:bandwidth}. Finally, in Section~\ref{sec:conclusion}, we conclude the paper with some open questions for future research.


\section{Preliminaries \label{sec:bg}}

In this section, we first present the formal definitions of convertible codes and the access cost associated with the conversion process. We then introduce the notion of \emph{per-symbol access optimality}. Finally, we define the bandwidth cost for the conversion process in the context of vector codes, along with some known results that will be useful for presenting our bandwidth-optimal construction in Section~\ref{sec:bandwidth}.


Consider an $[n,k]$ linear code $\mathcal{C}$ over $\mathbb{F}_q$. Let $\underline{c} = (c_1, \ldots, c_n) \in \mathcal{C}$ denote a codeword of $\mathcal{C}$. The \emph{minimum (Hamming) distance} of $\mathcal{C}$ is defined as
\[
d_\text{min}(\mathcal{C}) \triangleq \min_{\underline{c}, \underline{c}' \in \mathcal{C},\ \underline{c} \neq \underline{c}'} \left|\{i \in [n] \mid c_i \neq c_i'\}\right|.
\]
The code is said to be \emph{maximum distance separable (MDS)} if $d_\text{min}(\mathcal{C}) = n - k + 1$. In this paper, we focus on code conversions among MDS codes with different parameters.

\begin{note}[MDS Conjecture] \normalfont 
For $k \in \{1,n-1\}$, MDS codes can be constructed over any finite field.
The MDS conjecture~\cite{MDSConjecture} states that 
    $[n>k+1,k]$ MDS codes exist over $\mathbb{F}_q$ if and only if $q \ge n-1$, except when $q$ is a power of $2$ and $k \in \{3, n - 3\}$, in which case $q \ge n - 2$.    
\end{note}

Any $[n,k]$ linear code can be represented as the row space of a $k \times n$ matrix, known as the generator matrix $\mathbf{G}$ of the code. The generator matrix is called systematic if it has the form $\mathbf{G} = [\mathbf{I}_k~\mathbf{P}]$, where $\mathbf{P}$ is a $k \times (n-k)$ matrix. We refer to $\mathbf{P}$ as the \emph{parity matrix} (not to be confused with the parity-check matrix of a code). Moreover, a matrix $\mathbf{A}$ of size $a \times b$ is said to be \emph{superregular} if every square submatrix of $\mathbf{A}$ is nonsingular. It is well known that a code $\mathcal{C}$ is MDS if and only if it admits a systematic generator matrix $\mathbf{G} = [\mathbf{I}_k~\mathbf{P}]$, where $\mathbf{P}$ is superregular.

\subsection{Access-Optimality of Convertible Codes \label{sec:ao}}

Under the setting of code conversions between two MDS codes in the merge regime, we are given $\lambda \geq 2$ codewords of an $[n^I, k^I]$ initial MDS code $\mathcal{C}^I$, and the goal is to merge these codewords into a single codeword of an $[n^F, k^F = \lambda k^I]$ final MDS code $\mathcal{C}^F$. We assume that the $\lambda n^I$ symbols of the initial codewords are stored on distinct disks (or failure domains). Similarly, we assume that the $n^F$ symbols of the final codeword are stored on distinct disks. The formal definition of convertible codes in the merge regime is given below.

\begin{defn}[MDS Convertible Code]\normalfont
	An $(n^I,k^I;n^F,k^F)$ MDS convertible code is  defined by the three-tuple (i) an initial $[n^I,k^I]$ MDS code, (ii) a final $[n^F,k^F=\lambda k^I]$ MDS code, and (iii) a conversion procedure to convert $\lambda$ codewords of $\mathcal{C}^I$ into a codeword of $\mathcal{C}^F$.
\end{defn}


In~\cite{Maturana_Rashmi_TIT}, Maturana and Rashmi introduced the notion of \emph{access cost} to measure the efficiency of the conversion procedure, which consists of the following two aspects.

\textit{Write Access Cost:} When producing the final codeword, some symbols from the initial codewords may be reused directly. The \emph{write access cost} of a convertible code refers to the number of disks to which new symbols are written during the conversion. For an $[n^F,k^F]$ final code over $\mathbb{F}_q$, where symbols are stored on distinct disks, this is equivalently the number of newly generated symbols that are not reused from the initial codewords. It was shown in~\cite{Maturana_Rashmi_TIT} that the write access cost is at least $r^F = n^F - k^F$. This lower bound can be achieved by retaining $k^I$ symbols from each initial codeword in the final codeword. Throughout the paper, we assume this condition holds and thus $r^F$ new code symbols are written during the conversion process.

\textit{Read Access Cost:} The \emph{read access cost} of a convertible code refers to the number of disks from which data is read during the conversion. The default conversion scheme involves reading $k^I$ symbols from each of the $\lambda$ initial codewords and performing re-encoding to produce the final codeword. Since the symbols of initial codewords are assumed to be stored on distinct disks, this corresponds to accessing $\lambda k^I$ disks, resulting in a read access cost of $\lambda k^I$. It was shown in~\cite{Maturana_Rashmi_TIT} that if $r^F > \max\{k^I, r^I\}$ (where $r^I = n^I - k^I$), the read access cost is at least $\lambda k^I$, and thus the default scheme is optimal. However, if $r^F \leq \min\{k^I, r^I\}$, then code conversion can be performed by reading only $\lambda r^F$ disks. This motivates the following definition.




\begin{defn}[Access-Optimal MDS Convertible Code]\normalfont 
	Let $r^F\leq \min\{k^I,r^I\}$. An $(n^I,k^I;n^F,k^F)$ MDS convertible code is said to be access-optimal if the read access cost is $\lambda r^F$.
\end{defn}

 We  assume that $k^I>1$, as the $k^I=1$ case is trivial. It is also easy to see that $\lambda$ codewords of any initial $[n^I, k^I]$ MDS code in systematic form can be merged to obtain a codeword of a final $[n^F=\lambda k^I+1, k^F=\lambda k^I]$ MDS code by accessing the first parity symbol of the $\lambda$ initial codewords, resulting in a read access cost of $\lambda$. Therefore, the $r^F=1$ case is not of interest and we will restrict our attention to the setting $r^F>1$.


\subsection{Parity Matrix and Per-Symbol Access Optimality}\label{sec:pmatrix_persym_ao}

For an access-optimal MDS convertible code, the practical implication is that a coordinator node must download $\lambda r^F$ symbols from the initial codewords, generate all $r^F$ new symbols of the final codeword, and distribute them to $r^F$ distinct disks. We now introduce a stronger notion of access-optimality, referred to as \emph{per-symbol access-optimality}, wherein each new code symbol can be computed by reading exactly one symbol from each of the $\lambda$ initial codewords. This property is particularly beneficial in system settings without a central coordinator node, where new disks directly access the initial disks. While the standard access-optimality property guarantees optimality in terms of the aggregate read cost, it does not prevent individual new nodes from having to download more than $\lambda$ symbols in decentralized settings. In contrast, per-symbol access-optimality enables the conversion process to be performed in a fully parallelized manner, making it more suitable for such non-centralized architectures.




\begin{defn}[Per-Symbol Access-Optimal MDS Convertible Code]\label{def per-sym-opt}\normalfont Let $r^F\leq \min\{k^I,r^I\}$.
	An access-optimal $(n^I,k^I;n^F,k^F)$ MDS convertible code is said to be per-symbol access-optimal if each of the $r^F$ new symbols in the final codeword can be computed independently by reading exactly one symbol from each of the $\lambda$ initial codewords.
\end{defn}


Given the parameters ${n^I, k^I, k^F}$, if a per-symbol access-optimal MDS convertible code exists for $r^F = r^I$, then it can be trivially transformed into a per-symbol access-optimal code for any $r^F \leq r^I$, simply by omitting $r^I - r^F$ new symbols in the final codeword. Therefore, without loss of generality, we assume $r^F = r^I \triangleq r$ throughout this subsection and in Section~\ref{sec:sg_constr} which presents per-symbol access-optimal constructions.

Suppose the initial and final codes are defined by the systematic generator matrices $\mathbf{G}^I = [\mathbf{I}_{k^I}~\mathbf{P}^I]$ and $\mathbf{G}^F = [\mathbf{I}_{k^F}~\mathbf{P}^F]$, respectively. For each $\ell \in [\lambda]$, let $\underline{c}^{\ell} = (c^{\ell}_1, \ldots, c^{\ell}_{n^I})$ denote the $\ell$-th initial codeword. We regard the first $k^I$ symbols of each initial codeword as message symbols and the last $r$ symbols as parity symbols. Similarly, in the final codeword, the first $k^F$ symbols are considered message symbols and the last $r$ symbols are parity symbols. We assume that all $\lambda k^I$ message symbols from the initial codewords are retained unchanged in the final codeword, thereby ensuring write access optimality. In the following, we describe a sufficient condition on the matrix $\mathbf{P}^F$ that guarantees read access optimality. We assume $r \leq k^I$ since otherwise the default conversion approach already achieves read access optimality.

\begin{defn}[$(k^I,k^F,r)$-Parallel-Block-Reconstructible]\normalfont\label{def:parallel_rec}
	Let $\mathbf{P}$ be a $k^F \times r$ \emph{superregular} matrix, where $k^I \mid k^F$, and define $\lambda \triangleq \frac{k^F}{k^I}$. For each $\ell \in [\lambda]$, let $\mathbf{P}^{(\ell)}$ denote the restriction of $\mathbf{P}$ to the $k^I$ consecutive rows $[(\ell-1)k^I:\ell k^I-1]$, i.e.,
    \[
    \mathbf{P}^{(\ell)} \triangleq \mathbf{P}([(\ell-1)k^I : \ell k^I - 1], :).
    \]
    Decompose each $\mathbf{P}^{(\ell)}$ column-wise as
    \[
    \mathbf{P}^{(\ell)} = [\underline{p}^{(\ell,0)}\ \underline{p}^{(\ell,1)}\ \cdots\ \underline{p}^{(\ell,r-1)}],
    \]
    i.e., $\underline{p}^{(\ell,i)}$ is the $i$-th column of  $\mathbf{P}^{(\ell)}$. 
    Then, we say that $\mathbf{P}$ is \emph{$(k^I, k^F, r)$-parallel-block-reconstructible} if for every $\ell \in [\lambda]$ and every $i \in [0:r-1]$, there exists an index $i_{\ell} \in [0:r-1]$ such that $\underline{p}^{(\ell,i)}$ is a scalar multiple of $\underline{p}^{(1,i_{\ell})}$.
\end{defn}

\bnote \normalfont \label{remark:block_reconstr} We remark that the parallel-block-reconstructibility is a special case of block-reconstructibility defined in \cite{Maturana_Rashmi_TIT}. The matrix $\mathbf{P}$ is said to be block-reconstructible if for any $\ell \in [\lambda]$ each column of $\mathbf{P}^{(\ell)}$ lies in the column space of $\mathbf{P}^{(1)}$.   In \cite{Maturana_Rashmi_TIT}, block-reconstructible matrices are used to obtain access-optimal MDS convertible codes, whereas we obtain per-symbol access-optimal MDS convertible codes using parallel-block-reconstructible matrices. 
\enote

Suppose matrix $\mathbf{P}$ is $(k^I,k^F,r)$-parallel-block-reconstructible. If we set $\mathbf{P}^I=\mathbf{P}([0:k^I-1],:)$ and $\mathbf{P}^F=\mathbf{P}$, this will lead to an $(n^I=k^F+r,k^I;n^F=k^F+r,k^F)$ per-symbol access-optimal MDS convertible code. The proof follows simply from the following observation. Consider the $i$-th parity symbol of the final codeword, say ${d}_{k^F+i}$. The encoding vector corresponding to this parity symbol is given by (following the notation in Definition~\ref{def:parallel_rec}) 
$$[(\underline{p}^{(1,i)})^T\ (\underline{p}^{(2,i)})^T\ \cdots\ (\underline{p}^{(\lambda,i)})^T]^T.$$
Since $\underline{p}^{(\ell,i)}$ is a scalar multiple of $\underline{p}^{(1,i_{\ell})}$ for $\ell \in [\lambda]$, it follows that $d_{k^F+i}$ can be computed by reading exactly one parity symbol each from the $\lambda$ initial codewords. Specifically, $d_{k^F+i}$ can be computed by using $\{c^{\ell}_{k^I+i_{\ell}}\}_{\ell \in[\lambda]}$.
Thus we have the following result.

\blem\label{lem:per_sym_acc_opt} 
\normalfont Let $r \leq k^I$.
If the matrix $\mathbf{P}$ is $(k^I,k^F,r)$-parallel-block-reconstructible, then the systematic initial and final codes defined by the parity matrices $\mathbf{P}^I=\mathbf{P}([0:k^I-1],:)$ and $\mathbf{P}^F=\mathbf{P}$, respectively, yield an $(n^I=k^I+r,k^I;n^F=k^F+r,k^F)$ per-symbol access-optimal MDS convertible code. 
\elem 


\subsection{Bandwidth-Optimality of Convertible Codes} \label{sec:bandwidth_bg}
To optimize the bandwidth cost, Maturana and Rashmi~\cite{MaturanaRashmi_BWMerge_TIT} employ vector codes, wherein each code symbol is a vector over~$\mathbb{F}_q^\alpha$.
An $[n,k,\alpha]$ vector code encodes $k$ message symbols from $\mathbb{F}_q^\alpha$ into a codeword consisting of $n$ code symbols from $\mathbb{F}_q^\alpha$. The codes considered in the previous subsections on access-optimality are scalar codes, where each code symbol belongs to $\mathbb{F}_q$. A scalar code can be trivially viewed as a vector code with $\alpha = 1$. Similar to the scalar case, an $[n,k,\alpha]$ vector code is said to be a vector MDS code if any $k$ code symbols suffice to recover the original message. 

In the following context, when dealing with vector codes, we will often refer to the individual elements of a code symbol in $\mathbb{F}_q^\alpha$ as \emph{sub-symbols}, since each such element belongs to the base field $\mathbb{F}_q$. We use the notation $(n^I, k^I; n^F, k^F; \alpha)$-convertible code to denote a vector convertible code, where the initial and final codes are $[n^I, k^I, \alpha]$ and $[n^F, k^F, \alpha]$ vector MDS codes, respectively. In the merge regime, $\lambda \geq 2$ codewords of an initial $[n^I = k^I + r^I,\, k^I,\, \alpha]$ vector MDS code are merged into a single codeword of a final $[n^F = \lambda k^I + r^F,\, k^F = \lambda k^I,\, \alpha]$ vector MDS code. As in the case of scalar codes, we will continue to assume that the (vector) code symbols of the initial codewords --- and similarly, those of the final codeword --- are stored on distinct disks.

To define the bandwidth cost of conversion, it is necessary to fix a network model. The authors of~\cite{MaturanaRashmi_BWMerge_TIT} considered a model with a central coordinator node, which is also the model adopted in this paper. Disks involved in the conversion process are classified into three categories:
\begin{enumerate}
    \item \emph{Unchanged disks}, which store code symbols that are present in both the initial and final codewords without any modification;
    \item \emph{Retired disks}, which store code symbols of the initial codewords that are not present in the final codewords;
    \item \emph{New disks}, which store code symbols of the final codewords that are not present in the initial codewords.
\end{enumerate}

The \emph{read bandwidth} is defined as the total number of symbols over $\mathbb{F}_q$ (i.e., sub-symbols) transferred from the unchanged and retired disks to the coordinator node. The \emph{write bandwidth} is defined as the total number of symbols over $\mathbb{F}_q$ transferred from the coordinator node to the new disks. The \emph{bandwidth cost} is then the sum of the read and write bandwidths. In~\cite{MaturanaRashmi_BWMerge_TIT}, Maturana and Rashmi derived a lower bound on the bandwidth cost in the merge regime and provided a construction that matches this bound, thereby characterizing the optimal bandwidth cost given by~\eqref{eq:bandwidth}. In particular, the optimal write bandwidth is $r^F\alpha$, and the remaining part in~\eqref{eq:bandwidth} corresponds to the optimal read bandwidth.

 
\section{Sub-Group-Based Per-Symbol Access-Optimal Constructions \label{sec:sg_constr}}

In this section, we present two explicit constructions of 
MDS convertible code constructions that achieve per-symbol access-optimality. The constructions are based on $(k^I,k^F,r)$-parallel-block-reconstructible matrices, which as shown in Lemma~\ref{lem:per_sym_acc_opt}, guarantee per-symbol access-optimality. For simplicity and without loss of generality, we assume that $r^F=r^I=r$.

\subsection{Multiplicative Sub-Group-Based Construction}
\label{sec:constr1} 

To formally describe our construction, we begin by recalling the definition of a \emph{Cauchy matrix}.

\begin{defn}[Cauchy Matrix]\normalfont\label{def:Cauchy}
Let $m, n \in \mathbb{N}$. Consider subsets $X \triangleq \{x_i\}_{i=0}^{m-1}$ and $Y \triangleq \{y_j\}_{j=0}^{n-1}$ of $\mathbb{F}_q$ such that $X \cap Y = \emptyset$. The $m \times n$ matrix $\mathbf{P}$, with entries $\mathbf{P}(i,j) = (x_i - y_j)^{-1}$,
is called a \emph{Cauchy matrix}. Cauchy matrices are known to be superregular.
\end{defn}

Next, we first present a base construction that requires $r \mid (q-1)$ and $\lambda \leq r$. We will then introduce two minor variants of this construction later in the section to accommodate a broader range of parameters.

   \begin{constr} \label{constr:mult–subgrp} \normalfont 
The base construction is specified by a $k^F \times r$ Cauchy matrix $\mathbf{P}$, which is chosen to be $(k^I, k^F, r)$-parallel-block-reconstructible. For $i \in [0:k^F-1]$ and $j \in [0:r-1]$, the $(i,j)$-th entry of $\mathbf{P}$ is given by
\[
\mathbf{P}(i,j) = \frac{1}{x_i - y_j},
\]
where $x_i \in X$ and $y_j \in Y$, for two disjoint sets $X \triangleq \{x_i\}_{i=0}^{k^F-1}$ and $Y \triangleq \{y_j\}_{j=0}^{r-1}$ to be specified below. Let $\gamma$ be a primitive $r$-th root of unity. We set $y_j = \gamma^j$, so that $Y$ forms a multiplicative subgroup of $\mathbb{F}_q$. We construct $X$ as the disjoint union of $\lambda$ sets $\{X^{(\ell)}\}_{\ell=1}^\lambda$, each of which is disjoint from $Y$, has cardinality $k^I$, and satisfies the following property ({\bf P1}):

\textbf{P1:} For every $\ell \in[\lambda]$, there exists an $m_{\ell}\in [0:r-1]$ such that 
\[
X^{(\ell)} = \gamma^{m_{\ell}}X^{(1)}.
\]
   \end{constr} 
   We refer to the elements in $X^{(\ell)}$ as $\{x^{(\ell)}_t\}_{t=0}^{k^I - 1}$. Then, by the property {\bf P1}, we have $x^{(\ell)}_t = \gamma^{m_{\ell}} x^{(1)}_t$. Recall that $X$ is chosen as the disjoint union $\bigsqcup_{\ell=1}^{\lambda} X^{(\ell)}$. Thus we set
\[
x_{(\ell - 1)k^I + t} = x^{(\ell)}_t
\]
for $\ell \in [\lambda]$ and $t \in [0 : k^I - 1]$.


\begin{thm}\label{th:optimality_c1}
	Let matrix $\mathbf{P}$ be as described in Construction~\ref{constr:mult–subgrp}. Then $\mathbf{P}$ is $(k^I,k^F,r)$-parallel-block-reconstructible.
\end{thm}

\begin{IEEEproof}
	Fix $\ell \in[\lambda]$ and $j\in[0:r-1]$. Let
    \[
    \mathbf{P}^{(\ell)} = \mathbf{P}([(\ell-1)k^I:\ell k^I-1],:) = \left[ \underline{p}^{(\ell,0)}~\underline{p}^{(\ell,1)}~\cdots~\underline{p}^{(\ell,r-1)} \right]
    \]
    It suffices to show that $\underline{p}^{(\ell,j)}$ is a scalar multiple of $\underline{p}^{(1,j')}$, for some $j'\in[0:r-1]$. To prove this, consider the $(t,j)$-th entry of $\mathbf{P}^{(\ell)}$ for some $t\in[0:k^I-1]$ and $j \in [0:r-1]$. By the construction above, we have:
	$$\mathbf{P}^{(\ell)}(t,j)=\frac{1}{x_t^{(\ell)}-\gamma^j}=\frac{1}{\gamma^{m_\ell}x_t^{(1)}-\gamma^j}=\frac{\gamma^{-m_\ell}}{x_t^{(1)}-\gamma^{j'}},$$ 
    where $j'=(j-m_{\ell}) \mod r$. Thus, we conclude that $\underline{p}^{(\ell,j)}=\gamma^{-m_{\ell}}\underline{p}^{(1,j')}$.
\end{IEEEproof}

With the help of Theorem~\ref{th:optimality_c1}, we will now present an explicit choice of $\{X^{(\ell)}\}_{\ell=1}^\lambda$ satisfying property~{\bf P1}, which yields per-symbol access-optimal MDS convertible codes over a field size linear in the block length of the final code. Let $\alpha$ be a primitive element of $\mathbb{F}_q$, and define $\nu \triangleq \frac{q-1}{r}$. Set $\gamma = \alpha^\nu$. Choose $X^{(1)}$ as any subset of $\{\alpha, \alpha^2, \ldots, \alpha^{\nu-1}\}$, and define $X^{(\ell)} \triangleq \gamma^{\ell-1} X^{(1)}$ for $\ell \in [\lambda]$. For any parameters $k^I$, $r$, and $\lambda \le r$, this choice of $\{X^{(\ell)}\}_{\ell=1}^\lambda$ leads to the field size requirement $r \mid (q - 1)$ and $q \ge (k^I + 1)r + 1$. The restriction $\lambda \le r$  ensures that the sets $X^{(\ell)}$'s are disjoint.

Furthermore, we present the following two minor variants of Construction~\ref{constr:mult–subgrp}, which yield per-symbol access-optimal MDS convertible codes over even smaller fields.

\begin{itemize}
    \item[(i)] \textit{Construction 1-A}: The only difference from Construction~\ref{constr:mult–subgrp} is that we now include $0$ in the set $Y$. Observe that
    \[
    \frac{1}{x_t^{(\ell)} - 0} = \frac{1}{\gamma^{m_{\ell}} x_t^{(1)}} = \frac{1}{\gamma^{m_{\ell}} (x_t^{(1)} - 0)},
    \]
    which implies that the Cauchy matrix $\mathbf{P}$ remains parallel-block-reconstructible. The parameter constraint and field size requirement are now $\lambda \leq r - 1$ and $q$ such that $(r - 1) \mid (q - 1)$ and $q \ge (k^I + 1)(r - 1) + 1$.
    
    \item[(ii)] \textit{Construction 1-B}: In this variant, we append an all-one column to the Cauchy matrix $\mathbf{P}$ obtained in Construction 1-A. It is well known~\cite{RothSeroussiMDS} that the resulting matrix remains superregular. Moreover, the parallel-block-reconstructibility property is preserved as the appended column is an all-one column. The parameter constraint and field size requirement for this construction are $\lambda \leq r - 2$ and $q$ such that $(r - 2) \mid (q - 1)$ and $q \ge (k^I + 1)(r - 2) + 1$.
\end{itemize}

\begin{note}[Field Size Optimality]\normalfont
	Assume that $(r - 2) \mid (q - 1)$, $\lambda = r - 2$, and $k^I = \nu - 1$, where $\nu \triangleq \frac{q - 1}{r - 2}$. Let $\gamma = \alpha^{\nu}$, where $\alpha$ is a primitive element of $\mathbb{F}_q$. By choosing
    \[
    X^{(\ell)} = \gamma^{\ell - 1} \{ \alpha, \alpha^2, \ldots, \alpha^{\nu - 1} \}, \quad \text{for } \ell \in [1 : r - 2],
    \]
    in Construction 1-B, we obtain the final code parameters $n^F = \lambda k^I + r = q + 1$ and $k^F = \lambda k^I$. Assuming the MDS conjecture~\cite{MDSConjecture} holds, this code is field-size-optimal if $r>3$.
\end{note}

\begin{example}\normalfont\label{ex:1}
	Consider the prime field $\mathbb{F}_{13}$. Let $n^I = 9$, $k^I = 5$, $n^F = 14$, $k^F = 10$, and $\lambda = 2$. Therefore $r=4$ and $\nu =\frac{q-1}{r-2} = 6$. Note that $\alpha = 2$ is a primitive element. We choose $\gamma = 2^6 = 12 \pmod{13}$. Then, we have
    $Y = \{1, 12, 0\}$ and 
    \[X = \{2, 2^2, 2^3, 2^4, 2^5, 2^7, 2^8, 2^9, 2^{10}, 2^{11}\} = \{2, 4, 8, 3, 6, 11, 9, 5, 10, 7\}.
    \]
    The resultant $(5, 10, 4)$-parallel-block-reconstructible matrix $\mathbf{P}$ (after appending the all-one column as described in Construction 1-B) is illustrated in Fig.~\ref{fig:parity_mult_sg}.

	
\end{example}
\subsection{Additive Sub-Group-Based Construction}\label{sec:constr2}

Similar to Section~\ref{sec:constr1}, we first present a base construction that requires $r \mid q$ and $\lambda \leq r$, by specifying a $k^F \times r$ Cauchy matrix $\mathbf{P}$ that is $(k^I, k^F, r)$-parallel-block-reconstructible. Then, we introduce a minor variant of this base construction that has a reduced field size requirement.

\begin{constr} \normalfont \label{constr:add-subgrp}
    Following the notation in Section ~\ref{sec:constr1}, the Cauchy matrix $\mathbf{P}$ is described by two disjoint sets $X$ and $Y$. Analogous to Construction~\ref{constr:mult–subgrp}, here we  set $Y$ to be an additive subgroup of $\mathbb{F}_q$. The collection of  disjoint sets $\{X^{(\ell)}\}_{\ell =1}^\lambda$, each of size $k^I$, are chosen so that the following property is satisfied ({\bf P2}):

\textbf{P2:} For every $\ell \in[\lambda]$, there exists an $m_{\ell}\in [0:r-1]$ such that 
\[
X^{(\ell)} = y_{m_{\ell}}+X^{(1)}.
\]

\end{constr}

We refer to the elements in $X^{(\ell)}$ as $\{x^{(\ell)}_t\}_{t=0}^{k^I-1}$. Then, by the property \textbf{P2}, we have $x^{(\ell)}_t=y_{m_{\ell}}+x^{(1)}_t$. We choose $X$ as the disjoint union $\sqcup_{\ell=1}^{\lambda} X^{(\ell)}$. More specifically, for every $\ell\in[\lambda]$ and $t\in[0:k^I-1]$, we set $x_{(\ell-1)k^I+t}=x^{(\ell)}_t$.


\begin{thm}\label{th:optimality_c2}
	Let $\mathbf{P}$ be as described in Construction~\ref{constr:add-subgrp}. Then $\mathbf{P}$ is $(k^I,k^F,r)$-parallel-block-reconstructible.
\end{thm}

\begin{IEEEproof}
	The proof is analogous to that of Theorem~\ref{th:optimality_c1}. The key idea is that for any two elements $y_i, y_j \in Y$, we have $(y_i - y_j) \in Y$ since $Y$ is an additive subgroup. Fix $\ell \in [\lambda]$ and $j \in [0:r-1]$. We follow the notation of the column vector $\underline{p}^{(\ell,j)}$ introduced in the proof of Theorem~\ref{th:optimality_c1}. It suffices to show that $\underline{p}^{(\ell,j)}$ is a scalar multiple of $\underline{p}^{(1,j')}$ for some $j' \in [0:r-1]$. Consider the $t$-th entry (for $t \in [0:k^I - 1]$) of $\underline{p}^{(\ell,j)}$, which is defined as $\mathbf{P}^{(\ell)}( t, j)$. By the construction above, we have:
    $$
    \mathbf{P}^{(\ell)}(t, j) = \frac{1}{x_t^{(\ell)} - y_j} = \frac{1}{y_{m_{\ell}} + x_t^{(1)} - y_j} = \frac{1}{x_t^{(1)} - y_{j'}},
    $$
    where $y_{j'} = y_j - y_{m_{\ell}}$. Thus, we conclude that $\underline{p}^{(\ell,j)}=\underline{p}^{(1,j')}$.
\end{IEEEproof}
	

Next, we present an explicit choice for $\{X^{(\ell)}\}_{\ell=1}^\lambda$ satisfying property {\bf P2}, which yields per-symbol access-optimal MDS convertible codes by Theorem \ref{th:optimality_c2} with field size linear in $k^I$.

Let $p$ be a prime and $q=p^m$ for some positive integer $m$. Let $r=p^u$ for some $1\leq u <m$ . For clarity, we use the field isomorphism between $\mathbb{F}_q$ and the set of all polynomials in $\mathbb{F}_p[x]$ of degree $\leq (m-1)$. We choose $Y$ to be the set of all polynomials of degree $\leq (u-1)$ and $X^{(1)}$ as any subset of non-zero polynomials of the form $x^u f(x)$, where $\deg(f(x))\leq (m-u-1)$. Let $\{y_{m_\ell}\}_{\ell\in[\lambda]}$ denote $\lambda$ distinct elements in $Y$, with $y_{m_1}\triangleq 0$. We set $X^{(\ell)}=y_{m_\ell}+X^{(1)}$. For any $k^I, r$ and $\lambda \le r$, the field size requirement is that $r \mid q$ and $q \ge (k^I+1)r$.  


We have the following minor variant of construction II, which is analogous to Construction 1-B.
\begin{itemize}
    \item[(i)] \textit{Construction 2-A}: We append an all-one column to the Cauchy matrix $\mathbf{P}$ obtained in Construction~\ref{constr:add-subgrp}. The resulting matrix remains superregular, and the parallel-block-reconstructibility property is preserved. The parameter constraints for this construction are $\lambda \leq r - 1$ and $q$ such that $(r - 1) \mid q$ and $q \ge (k^I + 1)(r - 1)$.
\end{itemize}

\begin{note}[Field Size Optimality]\normalfont
	Assume that $(r-1)\mid q$, $\lambda = r-1$ and $k^I=p^{m-u}-1$. Using Construction 2-A, we have the final code parameters $n^F=\lambda k^I+r=q+1$ and $k^F=\lambda k^I$. Assuming that the MDS conjecture \cite{MDSConjecture} holds, this code is field-size-optimal if $r>3$.
\end{note}

\begin{example}\normalfont\label{ex:2}
	Let $n^I = 10$, $k^I = 7$, $n^F = 17$, $k^F = 14$, and $\lambda = 2$. Therefore $r=3$. We use the finite field $\mathbb{F}_{2^4}$ with primitive polynomial $x^4 + x + 1$ as the underlying field. The field elements, represented as polynomials of degree $\leq 3$, are indicated by their integer equivalents. For example, the polynomial $x^3 + x^2 + 1$, corresponding to the binary string $(1~1~0~1)$, is represented as $13$. Note that $\alpha = 2$ is a primitive element. We choose $Y = \{0, 1\}$ and 
    \[
    X = \{2, 4, 6, 8, 10, 12, 14, 3, 5, 7, 9, 11, 13, 15\}.
    \]
    The resulting $(7, 14, 3)$-parallel-block-reconstructible matrix $\mathbf{P}$ (after appending the all-one column as described in Construction 2-A) is illustrated in Fig.~\ref{fig:parity_additive_sg}.

\begin{figure}[htbp]
	\centering
	\begin{subfigure}[b]{0.45\columnwidth}
		\centering
		\includegraphics[scale=0.35]{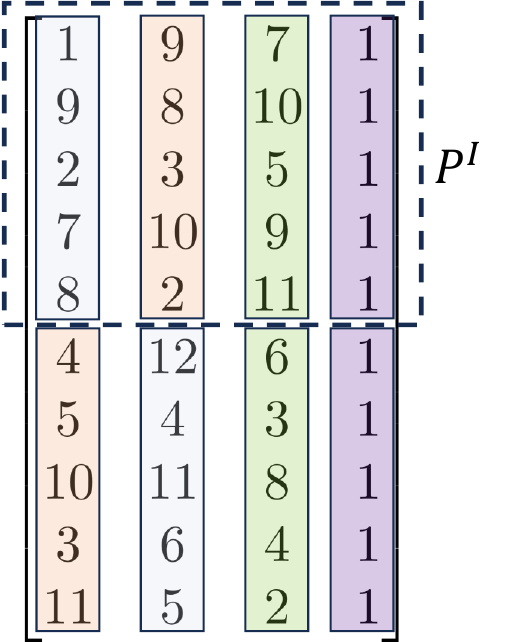}
		\caption{}
		\label{fig:parity_mult_sg}
	\end{subfigure}
	\hfill
	\begin{subfigure}[b]{0.45\columnwidth}
		\centering
		\includegraphics[scale=0.35]{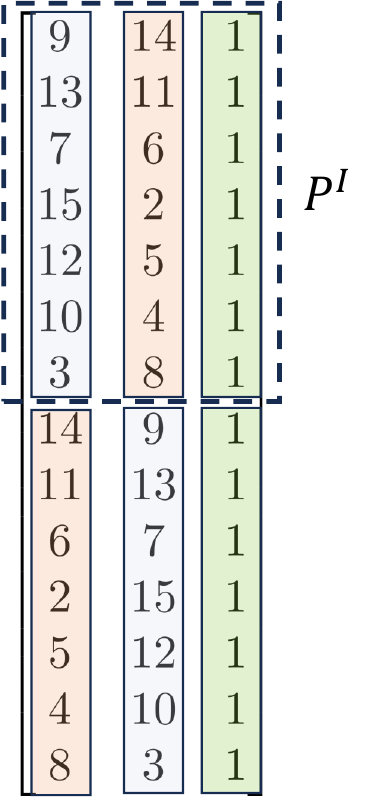}
		\caption{}
		\label{fig:parity_additive_sg}
	\end{subfigure}
	\caption{Illustration of the matrices $\mathbf{P}$ from Examples~\ref{ex:1} and~\ref{ex:2}. (a) Columns in the same color are scalar multiples of each other. The generator matrices corresponding to the initial and final codes are given by $[\mathbf{I}_5\ \mathbf{P}^I]$ and $[\mathbf{I}_{10}\ \mathbf{P}]$, respectively. (b) Columns in the same color are identical, corresponding to the generator matrices $[\mathbf{I}_7\ \mathbf{P}^I]$ and $[\mathbf{I}_{14}\ \mathbf{P}]$ for the initial and final codes, respectively.}
	\label{fig:combined_examples}
\end{figure}
\end{example}


\begin{note} \normalfont
While the initial conference version \cite{Convertible_ISIT25} of the current paper was under review, a pre-print \cite{ShiFangGao} appeared which also considers per-symbol access-optimality. The per-symbol access-optimal MDS convertible code constructions in \cite{ShiFangGao} are for $\lambda \le r$. This construction requires $q \ge \max \{n^F-1, n^I-1\}$, and additionally $r$ and $q$ must satisfy a specific constraint.
Since $\lambda \le r$ case is  covered by our constructions, \cite{ShiFangGao} does not provide a broader parameter range than our constructions. However, it is possible that the codes in \cite{ShiFangGao} require a slightly smaller field size than our constructions for some parameters.
\end{note}
\section{
Access-Optimal Constructions for All Parameters \label{sec:mod_poly_constr}}

In \cite{Kong}, Kong introduced an access-optimal $(n^I, k^I; n^F, k^F)$ MDS convertible code construction for the merge regime, where $k^F = \lambda k^I$, using a polynomial code framework. The construction is valid for all parameters $\{n^I, k^I, n^F, \lambda\}$, and it requires the underlying field size $q$ to satisfy $q \ge (\lambda+1)\max\{k^I, n^I - k^I\} + 1$ and $\max\{k^I, n^I - k^I\} \mid (q - 1)$.
In this section, working within the framework of parity-check matrices, we modify the polynomial-based construction from \cite{Kong} to obtain access-optimal MDS convertible codes under less stringent field size constraints. As a result, we show that access-optimal MDS convertible codes exist over fields as small as those allowed by the MDS conjecture in almost all cases.

As in previous sections, we restrict our attention to the case where $r^F \triangleq n^F - k^F \leq \min\{k^I, r^I\}$, since access optimality can otherwise be trivially achieved using the default conversion procedure. For simplicity, we omit the superscript on $k^I$ and henceforth refer to it simply as $k$ in this section. From here on, whenever we consider a subset of $\mathbb{F}_q$, we assume an implicit ordering, which will be clear from the context. For a subset $A = \{a_1, a_2, \ldots, a_k\} \subseteq \mathbb{F}_q$, we define the Vandermonde matrix
\[
\mathbf{V}_{A,r} \triangleq
\begin{bmatrix}
1 & 1 & \cdots & 1 \\
a_1 & a_2 & \cdots & a_k \\
\vdots & \vdots & \ddots & \vdots \\
a_1^{r-1} & a_2^{r-1} & \cdots & a_k^{r-1}
\end{bmatrix},
\]
which is an $r \times k$ matrix with $(i,j)$-th entry equal to $a_j^{i-1}$.
We further define
\[
\tilde{\mathbf{V}}_{A,{r}} \triangleq \left[\mathbf{V}_{A,r} ~ \underline{e}^{r} \right],
\]
where the column vector $\underline{e}^r$ denotes the $r$-th standard basis vector in $\mathbb{F}_q^r$, i.e., $\underline{e}^r = ( \underbrace{0, \ldots,0}_{(r-1)\text{ zeros}},1)^\top$.

\subsection{Modified Polynomial-based Access-optimal Constructions}

We begin with the simpler case where $r^F=r^I=r$, and then extend the construction to the general case.

\begin{constr}[Case $r^F=r^I=r$]\label{constr3} \normalfont Given positive integers $k$, $r \leq k$, and $\lambda \geq 2$.
Suppose there exist $\lambda + 1$ pairwise disjoint subsets $A_1, A_2, \ldots, A_{\lambda}$ and $B$ of $\mathbb{F}_q$, where $|A_{\ell}|=k$ and $|B|=r$, satisfying the following condition: For each $1 \leq \ell \leq \lambda$, there exists an invertible matrix $\mathbf{D}_{\ell}$ of order $r$ such that
\begin{equation}\label{eq_cond1_constr3}
\mathbf{H}_{A_{\ell}} = \mathbf{D}_{\ell} \mathbf{H}_{A_1},
\end{equation}
where, for brevity, we use $\mathbf{H}_{A_{\ell}}$ and $\mathbf{H}_{B}$ to denote the Vandermonde matrices $\mathbf{V}_{A_{\ell},r}$ and $\mathbf{V}_{B,r}$, respectively. Then we define initial code $\mathcal{C}^I$ as the code with parity-check matrix
\[
\left[ \mathbf{H}_{A_1} \ \mathbf{H}_{B} \right],
\]
and define the final code $\mathcal{C}^F$ as the code with parity-check matrix
\[
\left[ \mathbf{H}_{A_1} \ \mathbf{H}_{A_2} \ \cdots \ \mathbf{H}_{A_{\lambda}} \ \mathbf{H}_B \right].
\]
\end{constr}

\begin{thm}\label{thm_constr3}
    The initial-final code pair $(\mathcal{C}^{I},\mathcal{C}^{F})$ defined in Construction \ref{constr3} yields a $(k+r,k; \lambda k+r,\lambda k)$ access-optimal MDS convertible code. 
\end{thm}

\begin{IEEEproof}
Since $A_1,\ldots,A_{\lambda}$ and $B$ are pairwise disjoint subsets of $\mathbb{F}_q$, and the underlying parity-check matrices are all  Vandermonde matrices, it follows that $\mathcal{C}^{I}$ and $\mathcal{C}^{F}$ are MDS codes with parameters $[n^I=k+r,k]$ and $[n^F=\lambda k+r, \lambda k]$, respectively. Next, we show that any $\lambda$ codewords from $\mathcal{C}^{I}$ can be merged into a codeword of $\mathcal{C}^{F}$ in an access-optimal manner. 
    
    Suppose we are given $\lambda$ codewords $\underline{c}^1, \underline{c}^2, \ldots, \underline{c}^{\lambda}$ from $\mathcal{C}^{I}$, which are encoded from message vectors $\underline{m}^1, \underline{m}^2, \ldots, \underline{m}^{\lambda}$, respectively. In the following, we use the notation $\underline{c}|_{R}$ to denote the vector obtained by restricting the vector $\underline{c}$ to the coordinates indexed by the set $R$. We claim that the following word
    \begin{equation*}
    \underline{d} = \left(\underline{c}^1|_{[k]}, \underline{c}^2|_{[k]}, \ldots, \underline{c}^{\lambda}|_{[k]}, \underline{p} \right)
    \end{equation*}
    is a codeword in $\mathcal{C}^{F}$, where $\underline{p} = (p_1, \ldots, p_r)$ is defined as
    \begin{equation}\label{eq:p_defn}
    \underline{p}^{T} \triangleq 
    \left[
        \begin{array}{ccc}
            \mathbf{H}_{B}^{-1} \mathbf{D}_1 \mathbf{H}_{B} & \cdots & \mathbf{H}_{B}^{-1} \mathbf{D}_{\lambda} \mathbf{H}_{B}
        \end{array}
    \right]
    \left[
        \begin{array}{c}
            (\underline{c}^1|_{[k+1: n^I]})^{T} \\
            \vdots \\
            (\underline{c}^{\lambda}|_{[k+1: n^I]})^{T}
        \end{array}
    \right].
    \end{equation}

  By construction, the first $k$ symbols from each $\underline{c}^{\ell}$, i.e., symbols from $\underline{c}^{\ell}|_{[k]}$, are present unchanged in $\underline{d}$. Since each message vector $\underline{m}^{\ell}$ can be recovered from the corresponding $\underline{c}^{\ell}|_{[k]}$, it follows that $\underline{d}$ encodes the message $\left(\underline{m}^1, \underline{m}^2, \ldots, \underline{m}^{\lambda}\right)$. To construct the parity symbols of $\underline{d}$, it suffices to access only the parity symbols of $\underline{c}^1, \underline{c}^2, \ldots, \underline{c}^{\lambda}$. This provides an access-optimal conversion procedure from $\underline{c}^1, \underline{c}^2, \ldots, \underline{c}^{\lambda}$ to $\underline{d}$. We will now prove the claim that $\underline{d} \in \mathcal{C}^F$.

    Since $\underline{c}^{\ell} \in \mathcal{C}^I$, it follows that
    \begin{equation}\label{eq:initial_code_parity_check}
    \left[
    \begin{array}{cc}
   \mathbf{H}_{A_1} &  \mathbf{H}_B
    \end{array}
    \right]
    \begin{bmatrix}
    (\underline{c}^{\ell}|_{[k]})^T \\
    (\underline{c}^{\ell}|_{[k+1: n^I]})^T
    \end{bmatrix}
    = \underline{0}
    \end{equation}
    holds for each $1 \leq \ell \leq \lambda$, where $\underline{0}$ denotes the all-zero column vector. Consequently, as $\mathbf{H}_B$ is invertible, we have

    \begin{equation*}
    \left[
    \begin{array}{cc}
    \mathbf{H}_B^{-1} \mathbf{D}_\ell \mathbf{H}_{A_1} & \mathbf{H}_B^{-1} \mathbf{D}_\ell \mathbf{H}_B
    \end{array}
    \right]
    \begin{bmatrix}
    (\underline{c}^{\ell}|_{[k]})^T \\
    (\underline{c}^{\ell}|_{[k+1: n^I]})^T
    \end{bmatrix}
    = \underline{0}.
    \end{equation*}
    This implies
    \begin{equation*}
    (\mathbf{H}_B^{-1} \mathbf{D}_{\ell} \mathbf{H}_{A_1}) \cdot (\underline{c}^{\ell}|_{[k]})^T
    = -(\mathbf{H}_B^{-1} \mathbf{D}_{\ell} \mathbf{H}_B) \cdot (\underline{c}^{\ell}|_{[k+1: n^I]})^T.
    \end{equation*}
    By substituting this in the definition \eqref{eq:p_defn} of $\underline{p}$, we have
    \[
    \sum_{\ell=1}^{\lambda} (\mathbf{H}_B^{-1} \mathbf{D}_{\ell} \mathbf{H}_{A_1}) \cdot (\underline{c}^{\ell}|_{[k]})^T + \underline{p}^T = \underline{0}.
    \]
    Using condition~\eqref{eq_cond1_constr3}, this implies
    \[
    \sum_{\ell=1}^{\lambda} \mathbf{H}_{A_{\ell}} \cdot (\underline{c}^{\ell}|_{[k]})^T + \mathbf{H}_B \cdot \underline{p}^T = \underline{0},
    \]
    which confirms that $\underline{d} \in \mathcal{C}^F$. This completes the proof.
\end{IEEEproof}
Note that, to show that $\underline{d} = (\underline{c}^1|_{[k]}, \underline{c}^2|_{[k]}, \ldots, \underline{c}^{\lambda}|_{[k]}, \underline{p})$ lies in the null space of $\left[ \mathbf{H}_{A_1} \ \mathbf{H}_{A_2} \ \cdots \ \mathbf{H}_{A_{\lambda}} \ \mathbf{H}_B \right]$, the key equation used is \eqref{eq:initial_code_parity_check}, along with the condition \eqref{eq_cond1_constr3} and the definition \eqref{eq:p_defn}. Building on this observation, we can extend the arguments in the proof of Theorem~\ref{thm_constr3} to obtain the following result for the general case $r^F \leq \min\{k, r^I\}$.


\begin{thm}[General Case]\label{thm_constr3_general} Given positive integers $k$, $r^I$, and $r^F \leq \min\{k, r^I\}$, and $\lambda \geq 2$, suppose there exist $\lambda + 2$ subsets $A_1, A_2, \ldots, A_{\lambda}$ and $B^I, B^F$ of $\mathbb{F}_q$ satisfying the following conditions:
\begin{itemize}
    \item[(1)] $|A_{\ell}| = k$ for all $\ell \in [\lambda]$, $|B^I| = r^I$, and $|B^F| = r^F$.
    \item[(2)] The sets $A_1$ and $B^I$ are disjoint.
    \item[(3)] The sets $A_1, A_2, \ldots, A_{\lambda}$ and $B^F$ are mutually disjoint.
    \item[(4)] $B^F \subseteq B^I$.
    \item[(5)] For each $1 \leq \ell \leq \lambda$, there exists an invertible matrix $\mathbf{D}_{\ell}$ of order $r^F$ such that
    \begin{equation}\label{eq_cond1_general}
        \mathbf{V}_{A_{\ell}, r^F} = \mathbf{D}_{\ell} \mathbf{V}_{A_1, r^F}.
    \end{equation}
\end{itemize}

Let $A_1 = \{a_1, \dots, a_k\}$, $B^I = \{b_1, \dots, b_{r^I}\}$, and $B^F = \{b_1, \dots, b_{r^F}\}$. Let $f(x) = \prod_{\beta \in B^I \setminus B^F} (x - \beta)$, with $f(x) = 1$ when $B^I = B^F$. Define the scaling vector $\underline{v} = (v_\alpha)$ as
\[
v_{\alpha} = \begin{cases}
f(\alpha)^{-1} & \text{if } \alpha \in A_1 \cup B^F, \\
1 & \text{if } \alpha \in B^I \setminus B^F.
\end{cases}
\]
Then the initial code $\mathcal{C}^I$ with parity-check matrix
\[
\left[ \mathbf{V}_{A_1, r^I} \; \mathbf{V}_{B^I, r^I} \right] \cdot \mathrm{diag}(v_{\alpha} \mid \alpha \in A_1 \cup B^I)
\]
and the final code $\mathcal{C}^F$ with parity-check matrix
\[
\left[ \mathbf{V}_{A_1, r^F} \; \mathbf{V}_{A_2, r^F} \; \cdots \; \mathbf{V}_{A_{\lambda}, r^F} \; \mathbf{V}_{B^F, r^F} \right]
\]
together yield a $(k + r^I, k; \lambda k + r^F, \lambda k)$ access-optimal MDS convertible code.
\end{thm}

\begin{IEEEproof}
Let the $\lambda$ initial codewords be given by $\underline{c}^1, \underline{c}^2, \ldots, \underline{c}^{\lambda}$. Similar to the proof of Theorem \ref{thm_constr3}, we will argue that the word 
\begin{equation*}
    \underline{d} = \left(\underline{c}^1|_{[k]}, \underline{c}^2|_{[k]}, \ldots, \underline{c}^{\lambda}|_{[k]}, \tilde{\underline{p}} \right)
    \end{equation*}
    is a codeword in $\mathcal{C}^{F}$. Here $\tilde{\underline{p}}$ can be obtained as a function of $\{{c}^1|_{[k+1: k+r^F]},\ldots,{c}^\lambda|_{[k+1: k+r^F]}\}$, which leads to access optimality.

From the definition of the initial code, we have
\begin{equation}
\label{eq:pc_grs}
\left[ \begin{array}{cc}
    \mathbf{V}_{A_1, r^I} & \mathbf{V}_{B^I, r^I}
\end{array} \right]
\cdot \mathrm{diag}(v_{\alpha} \mid \alpha \in A_1 \cup B^I)
\cdot \left[ \begin{array}{c}
    (\underline{c}^{\ell} \big|_{[k]})^{T} \\
    (\underline{c}^{\ell} \big|_{[k+1:n^I]})^{T}
\end{array} \right]
= \underline{0}.
\end{equation}
We claim that the first $r^F$ parities of the initial codewords are related to the message symbols through the following equation
\begin{equation}
\label{eq:pc_grs_short}
\left[ \begin{array}{cc}
    \mathbf{V}_{A_1, r^F} & \mathbf{V}_{B^F, r^F}
\end{array} \right]
\cdot \left[ \begin{array}{c}
    (\underline{c}^{\ell} \big|_{[k]})^{T} \\
    (\underline{c}^{\ell} \big|_{[k+1:k+r^F]})^T
\end{array} \right]
= \underline{0}.
\end{equation}
Note that \eqref{eq:pc_grs_short} is analogous to \eqref{eq:initial_code_parity_check} in the proof of Theorem \ref{thm_constr3}.
Then, the argument that $\underline{d}\in\mathcal{C}^F$ follows by replacing the matrices $\mathbf{H}_{A_i}$ and $\mathbf{H}_B$, as well as the parity symbol vector $\underline{p}^{T}$ in the proof of Theorem~\ref{thm_constr3}, with $\mathbf{V}_{A_i, r^F}$, $\mathbf{V}_{B^F, r^F}$, and
\begin{equation*}
    \tilde{\underline{p}}^{T} \triangleq 
    \left[
        \begin{array}{ccc}
            \mathbf{V}_{B^F,r^F}^{-1} \mathbf{D}_1 \mathbf{V}_{B^F,r^F} & \cdots & \mathbf{V}_{B^F,r^F}^{-1} \mathbf{D}_{\lambda} \mathbf{V}_{B^F,r^F}
        \end{array}
    \right]
    \left[
        \begin{array}{c}
            (\underline{c}^1|_{[k+1: k+r^F]})^{T} \\
            \vdots \\
            (\underline{c}^{\lambda}|_{[k+1: k+r^F]})^{T}
        \end{array}
    \right].
\end{equation*}
To prove the claim, we now show that equation~\eqref{eq:pc_grs} implies equation~\eqref{eq:pc_grs_short}. Define an $r^F \times r^I$ matrix $\mathbf{F}$ as follows:
\begin{equation}
\label{eq:F_defn}
\mathbf{F} = 
\left[ \begin{array}{ccccccccc}
     f_0 & f_1 & f_2 & \cdots & f_{r^I-r^F} & 0 & \cdots & 0 \\
     0 & f_0 & f_1 & \cdots & f_{r^I-r^F-1} & f_{r^I-r^F} & \cdots & 0 \\
     \multicolumn{8}{c}{\ddots} \\
     0 & \cdots & \cdots & f_0 & \cdots & \cdots & \cdots & f_{r^I-r^F}
\end{array} \right]
\end{equation}
where $f_i$ are the coefficients of the polynomial $f(x) = \prod_{j=r^F+1}^{r^I} (x - b_j) = \sum_{i=0}^{r^I - r^F} f_i x^i$.

Since $v_{\alpha} = f(\alpha)^{-1}$ for $\alpha \in A_1 \cup B^F$, multiplying the parity-check matrix of $\mathcal{C}^I$ by $\mathbf{F}$ on the left yields
\begin{align*}
&\begin{bmatrix}
     f(a_1) & f(a_2) & \cdots & f(a_k) & f(b_1) & \cdots & f(b_{r^I}) \\
     a_1 f(a_1) & a_2 f(a_2) & \cdots & a_k f(a_k) & b_1 f(b_1) & \cdots & b_{r^I} f(b_{r^I}) \\
     \vdots & \vdots & \ddots & & \vdots & \ddots & \vdots \\
     a_1^{r^F-1} f(a_1) & a_2^{r^F-1} f(a_2) & \cdots & a_k^{r^F-1} f(a_k) & b_1^{r^F-1} f(b_1) & \cdots & b_{r^I}^{r^F-1} f(b_{r^I})
\end{bmatrix} \\
&\quad \cdot \operatorname{diag}(v_{a_1}, \cdots, v_{a_k}, v_{b_1}, \cdots, v_{b_{r^I}})
= \begin{bmatrix}
\mathbf{V}_{A_1, r^F} & \mathbf{V}_{B^F, r^F} & \mathbf{0}
\end{bmatrix}.
\end{align*}
Therefore, multiplying both sides of equation~\eqref{eq:pc_grs} on the left by $\mathbf{F}$ yields
\begin{align*}
\begin{bmatrix}
\mathbf{V}_{A_1, r^F} & \mathbf{V}_{B^F, r^F} & \mathbf{0}
\end{bmatrix}
\begin{bmatrix}
(\underline{c}^{\ell} \mid_{[1:k]})^{T} \\
(\underline{c}^{\ell} \mid_{[k+1:n^I]})^{T}
\end{bmatrix}
=\begin{bmatrix}
\mathbf{V}_{A_1, r^F} & \mathbf{V}_{B^F, r^F}
\end{bmatrix}
\begin{bmatrix}
(\underline{c}^{\ell} \mid_{[1:k]})^{T} \\
(\underline{c}^{\ell} \mid_{[k+1:k+r^F]})^{T}
\end{bmatrix}
= \underline{0}.
\end{align*}
This completes the proof.
\end{IEEEproof}

\begin{cor}\label{coro_constr3} \normalfont
Given positive integers $k$, $r^I$, and $r^F \leq \min\{k, r^I\}$, and $\lambda \geq 2$, let $q$ be a prime power satisfying $q \geq \max\{k + r^I, \lambda k + r^F\}$, and let $\gamma$ be a primitive element of $\mathbb{F}_q$. For $1 \leq \ell \leq \lambda$, define
\[
A_{\ell} = \big\{ \gamma^{(\ell-1)k}, \gamma^{(\ell-1)k + 1}, \ldots, \gamma^{\ell k - 1} \big\},
\quad
B^F = \big\{ 0, \gamma^{\lambda k}, \ldots, \gamma^{\lambda k + r^F - 2} \big\},
\]
and let $B^I$ be any subset of $\mathbb{F}_q \setminus A_1$ containing $B^F$ with $|B^I| = r^I$. Then, take the invertible matrix $\mathbf{D}_{\ell}$ in \eqref{eq_cond1_general} as
\begin{equation*}\label{eq_constr3_matrixD}
\mathbf{D}_{\ell} = \mathrm{diag}\big(1, \gamma^{(\ell-1)k}, \ldots, \gamma^{(\ell-1)k(r^F-1)}\big).
\end{equation*}
This yields an explicit construction of an access-optimal MDS convertible code with the parameters stated in Theorem~\ref{thm_constr3_general}.
\end{cor}


\subsection{Doubly-Extended GRS Codes: Further Field Size Reduction}

The following lemma provides a construction of MDS codes of length $n \le q+1$ over $\mathbb{F}_q$, which are referred to as doubly-extended GRS codes.


\begin{lem}[Doubly-Extended GRS Codes, {\cite[Ex.~5.2]{Roth06introduction}}]\label{doubly/triply_extended_GRS}
\normalfont Let $q \ge n-1$ be a prime power. Let $A=\{a_1, a_2, \ldots, a_{n-1}\}$ be a collection of $n-1$ distinct elements of $\mathbb{F}_q$.
Then, the $[n, n-r]$ linear code $\mathcal{C}$ over $\mathbb{F}_q$, defined by the parity-check matrix
    \begin{equation*}
      \tilde{\mathbf{V}}_{A,r} \cdot \mathrm{diag}(v_1, v_2, \ldots, v_{n})
    \end{equation*}
    is an MDS code, where $v_1, v_2, \ldots, v_{n}$ are nonzero elements of $\mathbb{F}_q$. This code $\mathcal{C}$ is called a \emph{doubly-extended GRS code}.
\end{lem}

Next, using Lemma~\ref{doubly/triply_extended_GRS}, we extend Theorem~\ref{thm_constr3_general} to the case where both the initial and final codes are doubly extended GRS codes.

\begin{thm}\label{thm_constr3_doubly_extended_general} Given positive integers $k$, $r^I$, and $r^F \leq \min\{k, r^I\}$, and $\lambda \geq 2$, suppose there exist $\lambda + 2$ subsets $A_1, A_2, \ldots, A_{\lambda}$ and $B^I, B^F$ of $\mathbb{F}_q$ satisfying conditions~(2)--(5) in Theorem~\ref{thm_constr3_general}, along with the following modified condition~(1):
\begin{itemize}  
    \item[(1)] $|A_{\ell}|=k$ for all $\ell \in [\lambda]$, $|B^I|=r^I-1$ and $|B^F|=r^F-1$. 
\end{itemize}  
Let $A_1 = \{a_1, \dots, a_k\}$, $B^F = \{b_1, \dots, b_{r^F-1}\}$, and $B^I = \{b_1, \dots, b_{r^I-1}\}$, and let $f(x) = \prod\limits_{\beta  \in B^I \setminus B^F} (x - \beta)$, with $f(x)=1$ when $B^I=B^F$. Define the scaling vector $\underline{v} = (v_\alpha)$ as
\[
v_{\alpha} = \begin{cases}
f(\alpha)^{-1} & \text{if } \alpha \in A_1 \cup B^F, \\
1 & \text{if } \alpha \in B^I \setminus B^F.
\end{cases}
\]
Then, the initial code $\mathcal{C}^I$ defined by the parity check matrix
\[
 \left[ \mathbf{V}_{A_1,r^I}~\tilde{\mathbf{V}}_{B^I,r^I} \right] \cdot \mathrm{diag}(v_{a_1},\cdots,v_{a_k},v_{b_1},\cdots,v_{b_{r^I-1}},1),
 \]
and the final code $\mathcal{C}^F$ defined by the parity-check matrix 
\[
\left[ \mathbf{V}_{A_1,r^{F}} ~ \mathbf{V}_{A_2,r^F} ~ \cdots ~ \mathbf{V}_{A_{\lambda},r^F} ~ \tilde{\mathbf{V}}_{B^F,r^F} \right]
\]
together yield a $(k + r^I, k; \lambda k + r^F, \lambda k)$ access-optimal MDS convertible code.  
\end{thm}
\begin{IEEEproof}
The proof follows similar lines to that of Theorem~\ref{thm_constr3_general}. Let $\mathbf{F}$ be the matrix defined in equation~\eqref{eq:F_defn}.  
Then, by left-multiplying the parity-check matrix of $\mathcal{C}^I$ with $F$, we obtain
\begin{align*}
&\left[
\begin{array}{ccccccc}
f(a_1) & \cdots & f(a_k) & f(b_1) & \cdots & f(b_{r^I-1}) & 0 \\
a_1 f(a_1) & \cdots & a_k f(a_k) & b_1 f(b_1) & \cdots & b_{r^I} f(b_{r^I-1}) & 0 \\
\vdots & \ddots & & \vdots & \ddots & \vdots & \vdots \\
a_1^{r^F-1} f(a_1) & \cdots & a_k^{r^F-1} f(a_k) & b_1^{r^F-1} f(b_1) & \cdots & b_{r^I}^{r^F-1} f(b_{r^I-1}) & 1 \\
\end{array}
\right] \\
&\quad  \cdot
\mathrm{diag}(v_{a_1}, \dots, v_{a_k}, v_{b_1}, \dots, v_{b_{r^I-1}}, 1)
=
\left[ V_{A_1, r^F} \;\; V_{B^F, r^F} \;\; \mathbf{0} \;\; \underline{e}_{r^F} \right].
\end{align*}
Thus, the code symbols of the initial codewords satisfy the following equation
\[
\begin{bmatrix}
\mathbf{V}_{A_1, r^F} & \mathbf{V}_{B^F, r^F} & \mathbf{0} & \underline{e}_{r^F}
\end{bmatrix}
\begin{bmatrix}
(\underline{c}^{\ell}\big|_{[1:k]})^{T} \\
(\underline{c}^{\ell}\big|_{[k+1:k+r^F-1]})^{T} \\
(\underline{c}^{\ell}\big|_{[k+r^F:k+r^I-1]})^{T} \\
\underline{c}^{\ell}(k+r^I)
\end{bmatrix}
= \underline{0}.
\]
Define $R=[k+1:k+r^F-1]\cup\{k+r^I\}$, this implies that
\[
\begin{bmatrix}
\mathbf{V}_{A_1, r^F} & \tilde{\mathbf{V}}_{B^F, r^F}
\end{bmatrix}
\begin{bmatrix}
(\underline{c}^{\ell}\big|_{[1:k]})^{T} \\
(\underline{c}^{\ell}\big|_{R})^{T}
\end{bmatrix}
= \underline{0}.
\]
Then, the result follows directly by replacing $\mathbf{V}_{B^F, r^F}$ and the parity symbol vector $\tilde{\underline{p}}^{T}$ in the proof of Theorem~\ref{thm_constr3_general} with $\tilde{\mathbf{V}}_{B^F, r^F}$ and
\[
\left[
    \begin{array}{ccc}
        \tilde{\mathbf{V}}_{B^F, r^F}^{-1} \mathbf{D}_1 \tilde{\mathbf{V}}_{B^F, r^F} & \cdots & \tilde{\mathbf{V}}_{B^F, r^F}^{-1} \mathbf{D}_{\lambda} \tilde{\mathbf{V}}_{B^F, r^F}
    \end{array}
\right]
\left[
    \begin{array}{c}
        (\underline{c}^1|_{R})^{T} \\
        \vdots \\
        (\underline{c}^{\lambda}|_{R})^{T}
    \end{array}
\right],
\]
respectively.
\end{IEEEproof}

\begin{cor}\label{coro_constr_dobly extended} \normalfont
Given positive integers $k$, $r^I$, and $r^F \leq \min\{k, r^I\}$, and $\lambda \geq 2$, let $q$ be a prime power satisfying $q \geq \max\{k + r^I, \lambda k + r^F\}-1$, and let $\gamma$ be a primitive element of $\mathbb{F}_q$. For $1 \leq \ell \leq \lambda$, define
\[
A_{\ell} = \big\{ \gamma^{(\ell-1)k}, \gamma^{(\ell-1)k + 1}, \ldots, \gamma^{\ell k - 1} \big\},
\quad
B^F = \big\{ 0, \gamma^{\lambda k}, \ldots, \gamma^{\lambda k + r^F - 3} \big\},
\]
and let $B^I$ be any subset of $\mathbb{F}_q \setminus A_1$ containing $B^F$ with $|B^I| = r^I-1$. Then, take the invertible matrix $\mathbf{D}_{\ell}$ as
\begin{equation*}\label{eq_constr3_matrixD}
\mathbf{D}_{\ell} = \mathrm{diag}\big(1, \gamma^{(\ell-1)k}, \ldots, \gamma^{(\ell-1)k(r^F-1)}\big).
\end{equation*}
This yields an explicit construction of an access-optimal MDS convertible code with the parameters stated in Theorem~\ref{thm_constr3_doubly_extended_general}.
\end{cor}

Assuming the MDS conjecture~\cite{MDSConjecture} holds, the construction described in Corollary~\ref{coro_constr_dobly extended} is field-size-optimal for almost all parameters. In particular, it is field-size optimal if $r^I,r^F \neq 3$.

\bnote\normalfont 
In Lemma~\ref{doubly/triply_extended_GRS}, if $r = 3$ and $q \ge n-2$ is a power of $2$, then an $[n, n-3]$ linear code $\mathcal{C}$ over $\mathbb{F}_q$, defined by the parity-check matrix
\begin{equation*}
  \left[\mathbf{V}_{A,r} ~ \underline{e}^2 ~ \underline{e}^3 \right] \cdot \mathrm{diag}(v_1, v_2, \ldots, v_{n}),
\end{equation*}
is also an MDS code (see \cite[Ex.~5.3]{Roth06introduction}). Such a code is referred to as a \emph{triply-extended GRS code}. The proof of Theorem~\ref{thm_constr3} can also be extended to the case where both the initial code $\mathcal{C}^{I}$ and the final code $\mathcal{C}^{F}$ are triply-extended GRS codes. This yields an access-optimal MDS convertible code with required field size $q \geq \lambda k + 1$ for $r^I=r^F=3$.
\enote

\subsection{Connections to the Polynomial-Based Construction in \cite{Kong}} 
We conclude this section by explaining the connections between Construction \ref{constr3} and the polynomial-based construction introduced in \cite{Kong}. 
For simplicity, we restrict our discussion to the case when $r^I=r^F=r$.
Let $\mathbb{F}_q^{<k}[x]$ denote the set of all polynomials over $\mathbb{F}_q$ with degree less than $k$. The polynomial-based construction of MDS convertible codes in \cite{Kong} is based on a mapping between polynomials in $\mathbb{F}_q^{<k}[x]$ and polynomials in $\mathbb{F}_q^{<\lambda k}$. Specifically, \cite{Kong} defined a map $\phi:(\mathbb{F}_q^{<k}[x])^{\lambda}\rightarrow \mathbb{F}_q^{<\lambda k}[x]$ as
\[
   \phi: (f_1,\ldots,f_{\lambda}) \longmapsto g,
\]
such that for evaluation point sets $A_{\ell}=\{a_{\ell,1},a_{\ell,2},\ldots,a_{\ell,k}\}$, $1\leq \ell \leq \lambda$, and $B=\{b_1,b_2,\ldots,b_r\}$ satisfying certain properties (see \cite[Corollary II.1]{Kong}), it holds that 
\[
g(a_{\ell,i})=\theta_{\ell,i}f_{\ell}(a_{1,i})
\]
and 
\[
g(b_{j})=\sum_{(\ell,i)\in [\lambda]\times [r] }\eta_{\ell,i,j}f_{\ell}(b_i)
\]
where absolute constants $\theta_{\ell,i}$ and $\eta_{\ell, i,j}$ only depends on the indices $(\ell, i)$ and $(\ell, i,j)$ respectively. This leads to a conversion process with optimal access cost between the $[k+r,k]$ RS code $\mathcal{C}^{I}$, defined as
\[
\mathcal{C}^{I}\triangleq\left\{ \left( f(a_{1,1}), \ldots, f(a_{1,k}), f(b_1) \dots, f(b_r) \right) \,\middle|\, f \in \mathbb{F}_q^{<k}[x]\right\},
\]
and the $[\lambda k+r,\lambda k]$ GRS code $\mathcal{C}^{F}$, defined as
\[
\mathcal{C}^{F}\triangleq\left\{ \left( \theta_{1,1}^{-1}g(a_{1,1}), \ldots, \theta_{\lambda,k}^{-1}g(a_{\lambda,k}), g(b_1), \ldots, g(b_r) \right) \,\middle|\, f \in \mathbb{F}_q^{<k}[x]\right\}.
\]

Since GRS codes can also be characterized by Vandermonde-like parity-check matrices, the conversion procedure introduced in Theorem~\ref{thm_constr3} can be viewed as a reinterpretation of the above mapping in the language of parity-check matrices. The main difference is that both the initial and final codes in Theorem~\ref{thm_constr3} (Construction~\ref{constr3}) are defined by Vandermonde-like parity-check matrices and therefore are GRS codes. This makes the conversion procedure simpler when viewed from the perspective of parity-check matrices. 

We note that our choices of $A_i$'s in Corollary \ref{coro_constr3} remove the divisibility constraint in \cite{Kong} that $\max\{k, r^{I}\}\mid (q-1)$. Moreover, by Corollary \ref{coro_constr_dobly extended}, one obtains access-optimal MDS convertible codes over a field of size $q \geq \max\{k + r^I, \lambda k + r^F\} - 1$, which improves upon the field size requirement in \cite{Kong}.

\section{A Bandwidth-Optimal Construction With Small Sub-Packetization} \label{sec:bandwidth}

While our earlier focus has been on constructing access-optimal scalar MDS convertible codes, we now turn to bandwidth-optimal vector MDS convertible codes. As discussed in Section~\ref{sec:bandwidth_bg}, the optimal bandwidth cost for vector MDS convertible codes in the merge regime has been characterized in~\cite{MaturanaRashmi_BWMerge_TIT} and is given by \eqref{eq:bandwidth}. This optimal bandwidth can be achieved using scalar MDS codes ($\alpha=1$) when either $r^I \geq r^F$ or $k^I \leq r^F$. However, for the case where $k^I > r^F > r^I$, the bandwidth-optimal construction in~\cite{MaturanaRashmi_BWMerge_TIT} requires sub-packetization $\alpha = r^F$. Our work focuses on the $k^I > r^F > r^I$ case, where we modify the existing construction to reduce the sub-packetization requirement to $\alpha = r^F/\gcd(r^F,r^I)$.
We also argue that $\alpha = r^F/\gcd(r^F,r^I)$ is the minimal possible sub-packetization under certain assumptions. 
To the best of our knowledge, this constitutes the first progress toward resolving an open problem posed in \cite{MaturanaRashmi_BWMerge_TIT} to determine the optimal value of $\alpha$. 
Furthermore, our construction requires a field size of only $q \geq n^F - 1$, representing an additional practical advantage. This field size improvement arises directly from the access-optimal constructions provided in the current paper. We remark that a similar field size reduction is possible for the bandwidth-optimal construction in \cite{MaturanaRashmi_BWMerge_TIT} as well.
 \subsection{Modified Piggybacking-Based Construction} \label{sec:bw_constr}
 
Assume that $k^I > r^F > r^I$. Here we present a bandwidth-optimal convertible code construction with a smaller sub-packetization level of 
$$\alpha = \frac{r^F}{g}, ~~\text{where}~~g\triangleq\text{gcd}(r^F, r^I).$$  
Our construction is based on the piggybacking framework, following the ideas presented in~\cite{MaturanaRashmi_BWMerge_TIT}. The piggybacking framework, originally introduced in~\cite{RashmiSR17_piggyback}, is a method for constructing vector codes using a scalar code as a base. The vector code is formed by taking multiple instances of the base code and adding specially designed functions of the data---referred to as \emph{piggybacks}---from one instance to another. In~\cite{RashmiSR17_piggyback}, the piggybacking technique was introduced and applied to design vector codes with reduced bandwidth for repairing an erased code symbol.

As in \cite{MaturanaRashmi_BWMerge_TIT}, our code construction is described in four parts: (1) the base scalar code in the piggybacking framework, (2) the initial $[n^I=k^I+r^I, k^I]$ vector MDS code $\mathcal{C}^I$ with sub-packetization $\alpha=\frac{r^F}{g}$, (3) the final $[n^F=\lambda k^I+r^F, k^F=\lambda k^I]$ vector MDS code $\mathcal{C}^F$ with sub-packetization $\alpha=\frac{r^F}{g}$, and (4) the conversion process between them.

 
\subsubsection{Base code} As the initial base code, we can use the initial code of any $(k^I+r^F, k^I; \lambda k^I+r^F, \lambda k^I)$ access-optimal scalar MDS convertible code (see Section~\ref{sec:pmatrix_persym_ao}) satisfying certain properties listed below.
Let $\mathcal{C}^{I'}$ be the initial $[k^I+r^F, k^I]$  MDS code and $\mathcal{C}^{F'}$ be the final $[n^F, k^F]$  MDS code of an access-optimal construction. 
Similar to \cite{MaturanaRashmi_BWMerge_TIT}, we assume that the access-optimal MDS convertible code satisfies the following properties:

\begin{itemize}
\item [(1)] the initial and final codes are linear and systematic, 
\item [(2)] the $r^F$ parity symbols of the merged final codeword can be computed from
the $r^F$ parity symbols of each of the $\lambda$ initial codewords. 
\end{itemize}

We remark that all known constructions of access-optimal MDS convertible codes, including those presented in this paper, meet these properties.
If the initial and final codes are linear and systematic, then property (2) follows from block reconstructibility (see Remark~\ref{remark:block_reconstr} for definition). 
Using access-optimal constructions in the current paper to obtain the initial base code,  bandwidth-optimal convertible codes with field size requirement of $q \ge n^F-1$ can be constructed.



Let $[\mathbf{I}~\mathbf{P}^I]$ be the systematic generator matrix of $\mathcal{C}^{I'}$, where $\mathbf{P}^I$ is an $k^I \times r^F$ matrix. Further, we decompose $\mathbf{P}^{I}$ column-wise as 
\[
\mathbf{P}^I=[\underline{p}^{0}~\underline{p}^{1}~\cdots~\underline{p}^{r^F-1}],
\]
i.e., $\underline{p}^{i}$ is the $i$-th column of $\mathbf{P}^I$. Let the systematic generator matrix of $\mathcal{C}^{F'}$ be $[\mathbf{I}~\mathbf{P}^F]$, where $\mathbf{P}^F$ is a $\lambda k^I \times r^F$ matrix. For $\ell \in[\lambda]$, define $\mathbf{P}^{(\ell)}$ to be the restriction of $\mathbf{P}^F$ to the $k^I$ consecutive rows $[(\ell-1)k^I:\ell k^I-1]$, i.e., 
\[
\mathbf{P}^{(\ell)}\triangleq \mathbf{P}^F([(\ell-1)k^I:\ell k^I-1],:).
\]
Let $\mathbf{P}^{(\ell)}$ be decomposed in terms of the columns as 
\[
\mathbf{P}^{(\ell)}=[\underline{p}^{(\ell,0)}\ \underline{p}^{(\ell,1)}\ \cdots\ \underline{p}^{(\ell,r^F-1)}].
\]



\subsubsection{Initial Code} The initial $[n^I=k^I+r^I, k^I, \alpha=\frac{r^F}{g}]$ vector MDS code $\mathcal{C}^I$ is constructed by applying the piggybacking 
framework to the initial base code $\mathcal{C}^{I'}$. 
Consider $\lambda$ message vectors $\underline{m}^{\ell} $, $\ell \in [\lambda]$, each of size $k^I \alpha$, where $\alpha = \frac{r^F}{g}$, i.e., $\underline{m}^{\ell} \in \mathbb{F}_q^{k^I \alpha}$.  
We write $\underline{m}^{\ell}  = (\underline{m}_0^{\ell} , \cdots, \underline{m}_{\alpha-1}^{\ell} )$, where each $\underline{m}_j^{\ell}  = (m_{0,j}^{\ell} , \cdots, m_{k^I-1,j}^{\ell} ) \in \mathbb{F}_q^{k^I}$. Let $\underline{c}^{\ell}  = (\underline{c}_0^{\ell} , \underline{c}_1^{\ell} , \cdots, \underline{c}_{n^I-1}^{\ell} )$ denote the $\ell$-th codeword of $\mathcal{C}^I$, where each $\underline{c}_i^{\ell}  = (c_{i,0}^{\ell} , c_{i,1}^{\ell} , \cdots, c_{i,\alpha-1}^{\ell} ) \in \mathbb{F}_q^{\alpha}$.


The codeword $\underline{c}^{\ell} $ encodes the message vector $\underline{m}^{\ell} $.  
We set $c_{i,j}^{\ell}  = m_{i,j}^{\ell} $ for every $i \in [0:k^I-1]$, $j \in [0:\alpha-1]$, and $\ell \in [\lambda]$ to ensure that the code $\mathcal{C}^I$ is systematic.

Let $\beta = \frac{r^I}{g}$. The  $r^I$ parity symbols (i.e., $r^I \alpha$ parity sub-symbols) of $\underline{\mathbf{c}}^{\ell} $ are defined as:

\bea
\label{eq:initialpcc}
c_{k^I+i, j}^{\ell}  = \begin{cases}
	\underline{m}_j^{\ell}  \cdot \underline{p}^{i} & 0 \le j \le \beta-1\\
	\underline{m}_j^{\ell} \cdot \underline{p}^{i} + \underline{m}_{i_1}^{\ell}  \cdot \underline{p}^{r^I + (\alpha-\beta)i_2+(j-\beta)}  & \beta \le j \le \alpha-1.
\end{cases}
\eea
for $i \in [0:r^I-1]$ and $j \in [0:\alpha-1]$, where $i_1=\lfloor \frac{i}{g}\rfloor$ and $i_2 \in [0: g-1]$ satisfies $i_2 \equiv i \pmod{g}$. Here the sub-symbols of the form $\underline{m}_{i_1}^{\ell}  \cdot \underline{p}^{r^I + (\alpha-\beta)i_2+(j-\beta)}$ are called the piggybacks. 
By varying $i \in [0:r^I-1]$ and $j \in [\beta: \alpha-1]$, it can be seen that the piggybacks are precisely $\{\underline{m}_v^{\ell}  \cdot \underline{p}^{u} \mid u \in [r^I:r^F-1], v \in [0:\beta-1]\}$.  


We now show that $\mathcal{C}^I$ is a vector MDS code using the standard decoding technique in piggybacking framework as follows. Consider an arbitrary subset $R \subseteq [n^I]$ of size $k^I$ and suppose we know the $k^I$ code symbols $\underline{c}_i^{\ell} $ with $i \in R$.
To prove that $\mathcal{C}^I$ is a vector MDS code, it suffices to show that the message $\underline{m}^{\ell}$ can be decoded using these code symbols. A codeword $\underline{c}^{\ell}$ can be thought of as an $(n^I \times \alpha)$ array as shown in Table~\ref{table:initial_codewords}. Let $\mathcal{C}^{I''}$ be the punctured code obtained by removing the last $(r^F-r^I)$ coordinates of $\mathcal{C}^{I'}$. 
It follows that $\mathcal{C}^{I''}$ is an $[n^I, k^I]$ MDS code. 
Observe that $(c_{0,j}^{\ell} , c_{1,j}^{\ell} , \cdots, c_{n^I-1,j}^{\ell} )$ is a codeword of $\mathcal{C}^{I''}$ for all $j \in [0:\beta-1]$. Due to the MDS property of $\mathcal{C}^{I''}$, we can decode $\underline{m}_j^{\ell} $ for all $j \in [0:\beta-1]$ from the known symbols.  After that, consider any $j \in [\beta:\alpha-1]$.  For $i \in [0:r^I-1]$, we have $\lfloor \frac{i}{g}\rfloor \le \beta-1$. Therefore, 
we can cancel out the piggybacks from $(c_{0,j}^{\ell} , c_{1,j}^{\ell} , \cdots, c_{n^I-1,j}^{\ell} )$ to obtain a codeword of $\mathcal{C}^{I''}$. Again, it follows from the MDS property of $\mathcal{C}^{I''}$ that we can decode $\underline{m}_j^{\ell} $ for all $j \in [\beta:\alpha-1]$. 




\subsubsection{Final Code} The final code $\mathcal{C}^F$ is a vector MDS code with parameters $[n^F = \lambda k^I + r^F,\ k^F = \lambda k^I,\ \alpha = \frac{r^F}{g}]$. Let $\underline{d} = (\underline{d}_0, \underline{d}_1, \ldots, \underline{d}_{n^F - 1})$ denote the merged codeword of $\mathcal{C}^F$, where each $\underline{d}_i = (d_{i,0}, d_{i,1}, \ldots, d_{i, \alpha-1}) \in \mathbb{F}_q^\alpha$. We assign ${d}_{(\ell-1)k^I + i, j} = \underline{m}_{i,j}^{\ell} $ for every $i \in [0: k^I - 1]$, $j \in [0:\alpha-1]$, and $\ell \in [\lambda]$, which ensures that the code $\mathcal{C}^F$ is systematic. To fully specify $\mathcal{C}^F$, it then suffices to define the remaining $r^F$ parity symbols (i.e., the final $r^F \alpha$ parity sub-symbols of $\underline{d}$) as functions of the complete message vector $\underline{m} = (\underline{m}^1, \underline{m}^2, \ldots, \underline{m}^\lambda) \in \mathbb{F}_q^{ \lambda k^I \alpha}$. For each $i \in [0: r^F - 1]$ and $j \in [0: \alpha - 1]$, we define:
\begin{equation*}
\label{eq:finalpcc}
d_{\lambda k^I + i, j} = \sum_{\ell = 1}^{\lambda} \underline{m}_j^{\ell}  \cdot \underline{p}^{(\ell, i)}.
\end{equation*}
It can be verified that, for every $j \in [0: \alpha - 1]$, the vector $(d_{0,j}, d_{1,j}, \ldots, d_{n^F - 1, j})$ forms a codeword of the $[n^F, k^F]$ MDS code $\mathcal{C}^{F'}$. Therefore, the code $\mathcal{C}^F$ is indeed a vector MDS code.



\begin{example} \label{eg3}\normalfont 
	Let $\lambda = 2, k^I=8, r^I=2, r^F=6$. Therefore,  $g=\text{gcd}(r^I, r^F) = 2$. The two initial codewords are shown in the following two tables. The number of columns here correspond to the sub-packetization level $\alpha=\frac{r^F}{g}=3$. 
    For the same set of parameters, the construction in \cite{MaturanaRashmi_BWMerge_TIT} has sub-packetization level of $6$. 
    The number of rows is equal to $n^I=k^I+r^I = 10$. Let $\mathbf{G}=[\mathbf{I}~\mathbf{P}^I]$ be the generator matrix of $\mathcal{C}^{I'}$ and $\mathbf{P}^I = [\underline{p}^0~\underline{p}^1~\cdots~\underline{p}^5]$. The initial code is formed by first placing $\alpha = 3$ codewords of the punctured code $\mathcal{C}^{I''}$ of the initial base code $\mathcal{C}^{I'}$ and then adding piggybacks, as shown in red below.

\begin{table}[ht!]
	\bean
	\begin{array}{||c|c|c||} \hline \hline
	m_{0,0}^1 &  m_{0,1}^1 & m_{0,2}^1\\ \hline
	m_{1,0}^1 &  m_{1,1}^1 & m_{1,2}^1\\ \hline
	m_{2,0}^1 &  m_{2,1}^1 & m_{2,2}^1\\ \hline
	m_{3,0}^1 &  m_{3,1}^1 & m_{3,2}^1\\ \hline
	m_{4,0}^1 &  m_{4,1}^1 & m_{4,2}^1\\ \hline
	m_{5,0}^1 &  m_{5,1}^1 & m_{5,2}^1\\ \hline
	m_{6,0}^1 &  m_{6,1}^1 & m_{6,2}^1\\ \hline
	m_{7,0}^1 &  m_{7,1}^1 & m_{7,2}^1\\ \hline
	\underline{m}_{0}^1 \cdot \underline{p}^{0} &  \underline{m}_{1}^1 \cdot \underline{p}^{0}+{\color{red}\underline{m}_{0}^1 \cdot \underline{p}^{2}}  & \underline{m}_{2}^1 \cdot \underline{p}^{0}+{\color{red}\underline{m}_{0}^1 \cdot \underline{p}^{3}} \\ \hline
    \underline{m}_{0}^1 \cdot \underline{p}^{1} &  \underline{m}_{1}^1 \cdot \underline{p}^{1}+{\color{red}\underline{m}_{0}^1 \cdot \underline{p}^{4}}  & \underline{m}_{2}^1 \cdot \underline{p}^{1}+{\color{red} \underline{m}_{0}^1 \cdot \underline{p}^{5}} \\ \hline \hline
	\end{array} \ \ \ \  
\begin{array}{||c|c|c||} \hline \hline
	m_{0,0}^2 &  m_{0,1}^2 & m_{0,2}^2\\ \hline
	m_{1,0}^2 &  m_{1,1}^2 & m_{1,2}^2\\ \hline
	m_{2,0}^2 &  m_{2,1}^2 & m_{2,2}^2\\ \hline
	m_{3,0}^2 &  m_{3,1}^2 & m_{3,2}^2\\ \hline
    m_{4,0}^2 &  m_{4,1}^2 & m_{4,2}^2\\ \hline
	m_{5,0}^2 &  m_{5,1}^2 & m_{5,2}^2\\ \hline
	m_{6,0}^2 &  m_{6,1}^2 & m_{6,2}^2\\ \hline
	m_{7,0}^2 &  m_{7,1}^2 & m_{7,2}^2\\ \hline
	\underline{m}_{0}^2 \cdot \underline{p}^{0} &  \underline{m}_{1}^2 \cdot \underline{p}^{0}+{\color{red}\underline{m}_{0}^2 \cdot \underline{p}^{2}}  & \underline{m}_{2}^2 \cdot \underline{p}^{0}+{\color{red}\underline{m}_{0}^2 \cdot \underline{p}^{3}} \\ \hline
    \underline{m}_{0}^2 \cdot \underline{p}^{1} &  \underline{m}_{1}^2 \cdot \underline{p}^{1}+{\color{red}\underline{m}_{0}^2 \cdot \underline{p}^{4}}  & \underline{m}_{2}^2 \cdot \underline{p}^{1}+{\color{red}\underline{m}_{0}^2 \cdot \underline{p}^{5}} \\ \hline \hline
\end{array}
	\eean
    \caption{The two tables above correspond to two initial codewords of $\mathcal{C}^I$ with $[n^I = 10, k^I=8, \alpha=3]$. The piggybacked symbols are indicated in red. Each column here excluding the piggybacks is a codeword of the punctured code $\mathcal{C}^{I''}$ used in the initial code construction. \label{table:initial_codewords}}
\end{table}
The final merged codeword is shown in Table~\ref{table:final_codeword}. Each column here is a codeword of $\mathcal{C}^{F'}$ with the generator matrix $[\mathbf{I}~\mathbf{P}^F]$ where 
\bean
\mathbf{P}^F = \left[ \begin{array}{cccc}
     \underline{p}^{(0,0)}&  \underline{p}^{(0,1)} &  \underline{p}^{(0,5)}\\
     \underline{p}^{(1,0)}&  \underline{p}^{(1,1)} &  \underline{p}^{(1,5)}
\end{array} \right],
\eean
is block-reconstructable from $\mathbf{P}^I$.
\begin{table}[ht!]
\centering
	\bean
\begin{array}{||c|c|c||} \hline \hline
m_{0,0}^1 &  m_{0,1}^1 & m_{0,2}^1\\ \hline
	m_{1,0}^1 &  m_{1,1}^1 & m_{1,2}^1\\ \hline
	m_{2,0}^1 &  m_{2,1}^1 & m_{2,2}^1\\ \hline
	m_{3,0}^1 &  m_{3,1}^1 & m_{3,2}^1\\ \hline
	m_{4,0}^1 &  m_{4,1}^1 & m_{4,2}^1\\ \hline
	m_{5,0}^1 &  m_{5,1}^1 & m_{5,2}^1\\ \hline
	m_{6,0}^1 &  m_{6,1}^1 & m_{6,2}^1\\ \hline
	m_{7,0}^1 &  m_{7,1}^1 & m_{7,2}^1\\ \hline
    m_{0,0}^2 &  m_{0,1}^2 & m_{0,2}^2\\ \hline
	m_{1,0}^2 &  m_{1,1}^2 & m_{1,2}^2\\ \hline
	m_{2,0}^2 &  m_{2,1}^2 & m_{2,2}^2\\ \hline
	m_{3,0}^2 &  m_{3,1}^2 & m_{3,2}^2\\ \hline
    m_{4,0}^2 &  m_{4,1}^2 & m_{4,2}^2\\ \hline
	m_{5,0}^2 &  m_{5,1}^2 & m_{5,2}^2\\ \hline
	m_{6,0}^2 &  m_{6,1}^2 & m_{6,2}^2\\ \hline
	m_{7,0}^2 &  m_{7,1}^2 & m_{7,2}^2\\ \hline
	\underline{m}_{0}^1 \cdot \underline{p}^{(1,0)}+\underline{m}_{0}^2 \cdot \underline{p}^{(2,0)} &  	\underline{m}_{1}^1 \cdot \underline{p}^{(1,0)}+\underline{m}_{1}^2 \cdot \underline{p}^{(2,0)}  &	\underline{m}_{2}^1 \cdot \underline{p}^{(1,0)}+\underline{m}_{2}^2 \cdot \underline{p}^{(2,0)} \\ \hline 
	\underline{m}_{0}^1 \cdot \underline{p}^{(1,1)}+\underline{m}_{0}^2 \cdot \underline{p}^{(2,1)} &  	\underline{m}_{1}^1 \cdot \underline{p}^{(1,1)}+\underline{m}_{1}^2 \cdot \underline{p}^{(2,1)} &	\underline{m}_{2}^1 \cdot \underline{p}^{(1,1)}+\underline{m}_{2}^2 \cdot \underline{p}^{(2,1)} \\ \hline
	\underline{m}_{0}^1 \cdot \underline{p}^{(1,2)}+\underline{m}_{0}^2 \cdot \underline{p}^{(2,2)} &  	\underline{m}_{1}^1 \cdot \underline{p}^{(1,2)}+\underline{m}_{1}^2 \cdot \underline{p}^{(2,2)} &	\underline{m}_{2}^1 \cdot \underline{p}^{(1,2)}+\underline{m}_{2}^2 \cdot \underline{p}^{(2,2)} \\ \hline 
    \underline{m}_{0}^1 \cdot \underline{p}^{(1,3)}+\underline{m}_{0}^2 \cdot \underline{p}^{(2,3)} &  	\underline{m}_{1}^1 \cdot \underline{p}^{(1,3)}+\underline{m}_{1}^2 \cdot \underline{p}^{(2,3)}  &	\underline{m}_{2}^1 \cdot \underline{p}^{(1,3)}+\underline{m}_{2}^2 \cdot \underline{p}^{(2,3)} \\ \hline
	\underline{m}_{0}^1 \cdot \underline{p}^{(1,4)}+\underline{m}_{0}^2 \cdot \underline{p}^{(2,4)} &  	\underline{m}_{1}^1 \cdot \underline{p}^{(1,4)}+\underline{m}_{1}^2 \cdot \underline{p}^{(2,4)} &	\underline{m}_{2}^1 \cdot \underline{p}^{(1,4)}+\underline{m}_{2}^2 \cdot \underline{p}^{(2,4)} \\ \hline 
    \underline{m}_{0}^1 \cdot \underline{p}^{(1,5)}+\underline{m}_{0}^2 \cdot \underline{p}^{(2,5)} &  	\underline{m}_{1}^1 \cdot \underline{p}^{(1,5)}+\underline{m}_{1}^2 \cdot \underline{p}^{(2,5)} &	\underline{m}_{2}^1 \cdot \underline{p}^{(1,5)}+\underline{m}_{2}^2 \cdot \underline{p}^{(2,5)} \\ \hline \hline
\end{array}
\eean
\caption{Merged final codeword of $\mathcal{C}^F$ with parameters $(n^F =22 , k^F=16, \alpha = 3)$ is shown above. Note that each column in the table is a codeword of $\mathcal{C}^{F'}$, the final code of the base convertible code used in the construction. \label{table:final_codeword}}
\end{table}

\end{example}

\subsubsection{Conversion Procedure} During the conversion process, 
the $\lambda k^I$ disks storing the message symbols of the initial codewords remain unchanged and the 
$\lambda r^I$ disks storing parity symbols of the initial codewords are retired. 
The parity symbols of the final merged codeword are stored on $r^F$ new disks and hence the write bandwidth is $r^F \alpha$. 
To provide a bandwidth-optimal conversion procedure, we now need to show that a read bandwidth of 
\begin{equation*}\label{opt_read_bw}
 \gamma_R = \lambda \alpha \left(r^I + \left(1 - \frac{r^I}{r^F}\right) k^I \right)  
\end{equation*}
sub-symbols is enough for the coordinator node to compute the parity symbols of the final merged codeword. Recall that for all $j \in [0:\alpha-1]$, $(d_{0,j}, {d}_{1,j}, \cdots, {d}_{n^F-1,j})$ is a codeword of $\mathcal{C}^{F'}$ with ${d}_{(\ell-1)k^I+i,j} = m_{i,j}^{\ell} $ for all $i \in [0: k^I-1]$ and $\ell \in [\lambda]$.  
It follows from Property (2) that $\{{d}_{k^F+i,j} \mid i \in [0:r^F-1]\}$ can be computed if the coordinator node knows $\{\underline{m}_j^{\ell}  \cdot \underline{p}^{i} \mid i \in [0:r^F-1], \ell \in [\lambda] \}$. In summary, if the coordinator node is able to obtain $\{\underline{m}_j^{\ell}  \cdot \underline{p}^{i} \mid i \in [0:r^F-1], j \in [0:\alpha-1], \ell \in [\lambda] \}$ using $\gamma_R$ downloaded sub-symbols, then we have a bandwidth-optimal conversion procedure. Before presenting the general conversion procedure for all parameter settings, we first describe the conversion procedure for the initial–final code pair given in Example~\ref{eg3}.

\begin{example} \normalfont Let $\lambda = 2$, $k = 8$, $r^I = 2$, and $r^F = 6$, and let the initial and final codes be as described in Example~\ref{eg3}. In the conversion process, the central coordinator node downloads:
\bit 
\item  $6$ parity sub-symbols from each of the two initial codewords (last two rows shown in the tables corresponding to initial codewords) given by 
\bean & \{c_{8+i, j}^{\ell}  \mid i \in [0:1], j \in [0:2],  \ell \in [2]\} \\ & = \{\underline{m}_0^{\ell}  \cdot \underline{p}^{0}, \underline{m}_1^{\ell}  \cdot \underline{p}^{0}+\underline{m}_0^{\ell}  \cdot p^{2}, \underline{m}_2^{\ell}  \cdot \underline{p}^{0}+\underline{m}_0^{\ell}  \cdot \underline{p}^{3}, \underline{m}_0^{\ell}  \cdot \underline{p}^{1}, \underline{m}_1^{\ell}  \cdot \underline{p}^{1}+\underline{m}_0^{\ell}  \cdot \underline{p}^{4}, \underline{m}_2^{\ell}  \cdot p^{1}+\underline{m}_0^{\ell}  \cdot \underline{p}^{5}  \mid \ell \in [2]\}
\eean 
\item 16 message sub-symbols from each of the two initial codewords 
$$\{\underline{m}_j^{\ell}  \mid j \in [2], \ell \in [2]\}=\{m_{i,j}^{\ell}  \mid i \in [0:7], j \in [1:2], \ell \in [2]\}.$$ 
These sub-symbols correspond to the the last two columns of the first eight rows in the initial codeword tables (Table~\ref{table:initial_codewords}).
\eit 
The total read bandwidth is therefore $6*2+16*2=44$, which is optimal. Next, we show that the coordinator can compute $\{\underline{m}_j^{\ell}  \cdot \underline{p}^{i} \mid i \in [0:5],\ j \in [0:2],\ \ell \in [2]\}$ from the downloaded sub-symbols, and can therefore carry out the conversion.

\bit 
\item Since the coordinator node has access to the message symbols $\{\underline{m}_j^{\ell}  \mid j \in [2],\ \ell \in [2]\}$, it can compute the sub-symbols $\underline{m}_j^{\ell}  \cdot \underline{p}^{i}$ for all $i \in [0:5]$, $j \in [2]$, and $\ell \in [2]$. The sub-symbols $\underline{m}_0^{\ell}  \cdot \underline{p}^{i}$ for $i \in [0:1]$ and $\ell \in [2]$ are directly available from the downloaded parity sub-symbols. It remains to show that the coordinator can compute the sub-symbols $\underline{m}_0^{\ell}  \cdot \underline{p}^{i}$ for $i \in [2:5]$ and $\ell \in [2]$. These sub-symbols correspond precisely to the piggybacks.
    
\item The coordinator node recovers the piggybacked sub-symbols $\underline{m}_0^{\ell}  \cdot \underline{p}^{2}$, $\underline{m}_0^{\ell}  \cdot \underline{p}^{3}$, $\underline{m}_0^{\ell}  \cdot \underline{p}^{4}$, and $\underline{m}_0^{\ell}  \cdot \underline{p}^{5}$, for $\ell\in[2]$, by canceling the corresponding interference terms $\underline{m}_1^{\ell}  \cdot \underline{p}^{0}$, $\underline{m}_2^{\ell}  \cdot \underline{p}^{0}$, $\underline{m}_1^{\ell}  \cdot \underline{p}^{1}$, and $\underline{m}_2^{\ell}  \cdot \underline{p}^{1}$ from the downloaded parity sub-symbols $c_{8,1}^{\ell} $, $c_{8,2}^{\ell} $, $c_{9,1}^{\ell} $, and $c_{9,2}^{\ell} $, respectively. 
 \eit 
\end{example}

We now present the general conversion procedure, which follows a similar structure to the conversion process illustrated in the example above.

\ben
\item The coordinator node downloads the following $\lambda r^I \alpha$ parity sub-symbols and $\lambda(\alpha-\beta) k^I$ message sub-symbols: 
\bean
\{ c_{k^I+i, j}^{\ell}  \mid  i \in [0:r^I-1], j \in [0: \alpha-1], \ell \in [\lambda] \} \cup \{ m_{i, j}^{\ell}  \mid i \in [0:k^I-1], j \in [\beta: \alpha-1], \ell \in [\lambda] \}. 
\eean
This results in a total read bandwidth of $\lambda (r^I \alpha + (\alpha - \beta) k^I ) = \lambda \alpha \left(r^I + \left(1 - \frac{r^I}{r^F}\right) k^I \right)$, which meets the bound. Consequently, if the coordinator node is able to compute the set $\{\underline{m}_j^{\ell}  \cdot \underline{p}^{i} \mid i \in [0:r^F-1],\ j \in [0:\alpha-1],\ \ell \in [\lambda]\}$ from the downloaded sub-symbols, then by Property (2), it can obtain the contents of the new disks and complete the conversion procedure successfully. We note that $\underline{m}_j^{\ell}  \cdot \underline{p}^{i}=c_{k^I+i, j}^{\ell} $ have already been downloaded, for all $i \in [0:r^I-1], j \in [0:\beta-1]$ and $\ell \in [\lambda]$.


\item 
The coordinator node then computes sub-symbols $\{\underline{m}_j^{\ell}  \cdot \underline{p}^{i} \mid i \in [0:r^F-1], j \in[\beta:\alpha-1], \ell \in [\lambda]\}$ as it has access to message symbols $\underline{m}_j^{\ell} $ for all $j \in [\beta:\alpha-1]$ and $\ell \in [\lambda]$.

\item The remaining sub-symbols to be computed are $\underline{m}_j^{\ell}  \cdot \underline{p}^{i}$ for $i \in [r^I:r^F-1], j \in[0:\beta-1]$, and $\ell \in [\lambda]$, which are all the piggybacks. Note that the coordinator node has access to parity sub-symbols 
$$\{c_{k^I+i, j}^{\ell}  \mid  i \in [0:r^I-1], j \in [\beta:\alpha-1],\ell\in [\lambda]\}.$$ 
Thus, for each $j \in [\beta:\alpha-1]$, by~\eqref{eq:initialpcc}, the coordinator node can recover the piggyback $\underline{m}_{i_1}^{\ell}  \cdot \underline{p}^{r^I + (\alpha-\beta)i_2+(j-\beta)}$ from $c_{k^I+i, j}^{\ell} $ as it also has access to $\underline{m}_j^{\ell} $. By repeating this process for all $i \in [0:r^I-1]$, $j \in [\beta, \alpha-1]$ and $\ell \in [\lambda]$, all the piggybacks can be recovered.   
\een


\subsection{Optimality of Sub-Packetization Level} 
The sub-packetization level of $\alpha=\frac{r^F}{\gcd(r^F,r^I)}$ is the minimum possible for bandwidth-optimal MDS convertible codes under the assumption that the coordinator node downloads equal amount of data from every unchanged disk and the entire content of every retired disk. We note that both the construction in \cite{MaturanaRashmi_BWMerge_TIT} and our modified version in this paper satisfies this assumption.
It follows from \cite[Thm.~4]{MaturanaRashmi_BWMerge_TIT} that for bandwidth-optimal conversion, the number of unchanged disks is $\lambda k^I$ and the number of retired disks $\lambda r^I$. Furthermore, it can be inferred from \cite{MaturanaRashmi_BWMerge_TIT} that any bandwidth-optimal MDS convertible code must also be read-bandwidth-optimal.
For the parameter regime of interest $k^I > r^F > r^I$, the minimum read bandwidth is given by
$\gamma_R=\lambda r^I \alpha+ \lambda k^I \alpha (1-\frac{r^I}{r^F}).$ 
Therefore, if the assumption holds, then to achieve bandwidth-optimality, each unchanged disk must transmit $\alpha (1-\frac{r^I}{r^F})$ symbols to the coordinator node. This quantity should be an integer, i.e., $r^F$ must divide $\alpha r^I$. The smallest $\alpha$ for which this is true is $\frac{r^F}{\gcd(r^F,r^I)}$.   
\subsection{Bandwidth-Optimal Conversion for Multiple Values of $r^F$} 
The final code parameters might not be decided at the time of initial encoding. Therefore, the ability to simultaneously support multiple values of final parameters is an important practical consideration. 
The authors of \cite{MaturanaRashmi_BWMerge_TIT} present a construction  which is bandwidth-optimal  simultaneously for multiple values of $r_F \in \{r_1,\dots,r_s\}$ such that $r_i > r^I$.  Let $r=\max_{i \in [s]} \{r_i\}$. 
The construction in \cite{MaturanaRashmi_BWMerge_TIT}  uses their fixed $r^F$ construction as a building block. They require a $(k^I+r, k^I, \lambda k^I+r, \lambda k^I)$ access-optimal MDS convertible code with the following property in addition to the two properties in Section~\ref{sec:bw_constr}:  
\bit 
\item $\lambda$ codewords of initial $[k^I+r, k^I]$  MDS code can be merged in an access-optimal fashion to obtain a codeword of a $[\lambda k^I+r', \lambda k^I]$ MDS code for all $r' \in [r]$. 
 \eit 
We remark that all known constructions of access-optimal MDS convertible codes, including those presented in this paper, have this additional property as well. 
The sub-packetization requirement of the construction in \cite{MaturanaRashmi_BWMerge_TIT} is $\alpha= \prod_{i=1}^{s}r_i$. Using our fixed $r^F$ construction in Section~\ref{sec:bw_constr} as a building block, we can obtain an MDS convertible code construction which is bandwidth-optimal  simultaneously for multiple values of $r_F \in \{r_1,\dots,r_s\}$ such that $r_i > r^I$, with a smaller sub-packetization level of $\alpha= \prod_{i=1}^{s} \frac{r_i}{\gcd(r^I,r_i)}$. We skip the proof as it follows from the ideas presented in \cite{MaturanaRashmi_BWMerge_TIT}. 

\section{Conclusion}\label{sec:conclusion}

In this paper, we presented constructions of access-optimal and bandwidth-optimal MDS convertible codes in the merge regime. Under the access cost metric, we proposed a construction that achieves optimal access cost, applicable to all parameter regimes. The field size requirement of this construction matches the smallest possible field size permitted by the MDS conjecture for almost all parameters. Additionally, we introduced the notion of per-symbol access-optimality and provided low field size constructions for certain parameter settings. Nevertheless, the problem of constructing per-symbol access-optimal MDS convertible codes with low field size for general parameters remains open.

With respect to the bandwidth cost metric, we refined a previously known construction to reduce the sub-packetization level while preserving bandwidth optimality. Under certain natural assumptions, our construction achieves the optimal sub-packetization level. However, in the general case, determining the minimum sub-packetization level required for bandwidth-optimal MDS convertible codes without any assumptions remains an open problem.

\bibliographystyle{IEEEtran}
\bibliography{convcodes.bib}

\end{document}